\documentclass[12pt]{article}
\pdfoutput=1

\usepackage{cite}
\usepackage{setspace}
\usepackage{xspace}
\usepackage{color}
\usepackage{amsmath, amsfonts, amssymb, amsthm, upgreek, amstext}
\usepackage{caption}
\usepackage{slashed}
\usepackage{subfig}
\usepackage{multirow}
\usepackage{array} 
\usepackage{bbold}

\usepackage[english]{babel}
\usepackage[applemac]{inputenc}
\usepackage[colorlinks=true,linkcolor=black,anchorcolor=black,citecolor=blue,filecolor=cyan,menucolor=red,runcolor=filecolor,urlcolor=blue,bookmarks=true,bookmarksnumbered=true]{hyperref}
\usepackage[dvipsnames]{xcolor}

\usepackage{ifpdf}
\ifpdf
\usepackage[pdftex]{graphicx}
\usepackage{epstopdf}
\else
\usepackage[dvips]{graphicx}
\fi

\newcommand{\beq}{\begin{equation}}
\newcommand{\eeq}{\end{equation}}
\newcommand{\bea}{\begin{eqnarray}}
\newcommand{\eea}{\end{eqnarray}}
\newcommand{\bag}{\begin{align}}
\newcommand{\eag}{\end{align}}

\newcommand{\Eq}[1]{Eq.\!~(\ref{#1})}
\newcommand{\Fig}[1]{Fig.\!~\ref{#1}}

\newcommand{\Eqss}[3]{Eqs.\!~(\ref{#1}, \ref{#2}, \ref{#3})}

\newcommand{\MeV}{\,\mathrm{MeV}}
\newcommand{\GeV}{\,\mathrm{GeV}}
\newcommand{\TeV}{\,\mathrm{TeV}}

\newcommand{\fb}{\,\mathrm{fb}}
\newcommand{\ab}{\,\mathrm{ab}}

\newcommand{\nn}{\nonumber}

\newcommand{\ie}{$\textnormal{i.e.}$ }

\newcommand{\Tr}{\mathrm{Tr}}

\newcommand{\BR}{\mathrm{BR}}

\newcommand{\vev}[1]{\langle {#1} \rangle}

\newcommand{\hc}{\mathrm{h.c.}}

\newcommand{\Lt}{\widetilde L}
\newcommand{\Ct}{\widetilde C}
\newcommand{\Xt}{\widetilde X}
\newcommand{\Ht}{\widetilde H}
\newcommand{\Psit}{\widetilde \Psi}
\newcommand{\psit}{\tilde \psi}
\newcommand{\Wt}{\widetilde W}
\newcommand{\gt}{\tilde g}
\newcommand{\Z}{\widehat R}
\newcommand{\Y}{\widehat Y}
\newcommand{\Ah}{\widehat B}
\newcommand{\gh}{\hat g}
\newcommand{\thetat}{\tilde \theta}
\newcommand{\wt}{\tilde \omega}
\newcommand{\topt}{\tilde t}
\newcommand{\bt}{\tilde b}
\newcommand{\qt}{\tilde q}
\newcommand{\taut}{\tilde \tau}
\newcommand{\Qt}{\widetilde Q}
\newcommand{\vt}{\tilde v}
\newcommand{\yt}{\tilde y}
\newcommand{\Lambdat}{\Lambda_{\slashed{Z_2}}}
\newcommand{\mt}{\widetilde m}
\newcommand{\Lqcd}{\Lambda_{\mathrm{QCD}}}
\newcommand{\Lqcdt}{\tilde \Lambda_{\mathrm{QCD}}}

\newcommand{\tH}{twin-Higgs }

\begin{document}

\title{
\vspace{-1.5cm}
\normalsize{\hfill TUM-HEP-1199/19}\\
\normalsize{\hfill CERN-TH-2019-064}\\
\vspace{1.5cm}
\vspace{0.0 cm}
{\huge \bf Hypercharged Naturalness}}

\author{
Javi Serra$^a$, Stefan Stelzl$^a$, Riccardo Torre$^{b,c}$, and Andreas Weiler$^a$\\ \\
{\small\emph{$^a$ Physik-Department, Technische Universit\"at M\"unchen, 85748 Garching, Germany}}\\
{\small\emph{$^b$ Theoretical Physics Department, CERN, 1211 Geneva 23, Switzerland}}\\
{\small\emph{$^c$ INFN, Sezione di Genova, Via Dodecaneso 33, 16146 Genova, Italy}}\\
}
\date{}

\maketitle
\begin{abstract}
\noindent We present an exceptional twin-Higgs model with the minimal symmetry structure for an exact implementation of twin parity along with custodial symmetry. Twin particles are mirrors of the Standard Model yet they carry hypercharge, while the photon is identified with its twin. We thoroughly explore the phenomenological signatures of hypercharged naturalness: long-lived charged particles, a colorless twin top with electric charge $2/3$ that once pair-produced, bounds via twin-color interactions and can annihilate to dileptons or a Higgs plus a photon or a $Z$, and glueballs produced from Higgs decays and twin-quarkonium annihilation that either decay displaced, or are stable on collider scales and eventually decay to diphotons. Prospects for detection of these signatures are also discussed.

\end{abstract}

\newpage

\begingroup
\tableofcontents
\endgroup 

\setcounter{equation}{0}
\setcounter{footnote}{0}



\section{Introduction} \label{intro}

In an era of active experimental research but no direct signs of physics beyond the Standard Model, especially of the dynamics behind the insensitivity of the scalar Higgs field to higher mass scales, solutions to the electroweak hierarchy problem with non-standard low-energy signatures are called-for.

Recent progress in this direction ranges from theories where the gauge quantum numbers of the BSM states cutting off the UV contributions to the Higgs potential are unconventional, the prime realization of this idea being the so-called twin Higgs \cite{Chacko:2005pe,Barbieri:2005ri,Chacko:2005vw,Chacko:2005un}, to scenarios relying on a non-trivial cosmological history to explain the preference for a light Higgs sector \cite{Graham:2015cka,Arkani-Hamed:2016rle}.
The former alternative, which we focus on in this work, is easy to understand and to motivate: the prevailing exploration of the high-energy territory is due to the LHC, a hadron machine where colored BSM particles can be produced with large cross sections. This collider has already shown its constraining power on such type of new physics, e.g.~colored top partners in composite-Higgs models (see \cite{Contino:2010rs,Bellazzini:2014yua,Panico:2015jxa} for reviews), whose mass should now exceed a TeV \cite{Aaboud:2018pii,Sirunyan:2018qau}, and similar bounds apply to colored supersymmetric particles. While this might indicate that the characteristic scale of the new dynamics is higher than expected from naturalness considerations, there is also the possibility that nature has just chosen none of the simpler low-energy manifestations of the TeV solutions to the hierarchy problem, specifically those with light colored states.

Indeed, in composite-\tH models, the Higgs arises as a Nambu-Goldstone boson (NGB) from the spontaneous breaking of a global symmetry at a scale $f \approx \TeV$, as in standard theories of Higgs compositeness, yet crucially the leading radiative contributions to the Higgs potential, from loops of SM fields, are cut off by uncolored states, their twins, owing to a $Z_2$ symmetry that relates them. 
In particular, the large explicit breaking of the Higgs shift-symmetry from the top Yukawa coupling no longer results in a squared Higgs mass proportional to $(y_t/ 4\pi)^2 \, m_*^2$, with $m_* = g_* f$ the typical mass of the colored composite resonances, because a color-neutral twin top introduces a complementary breaking of the shift-symmetry, enforced by twin parity and resulting in a $(y_t/g_*)^2$ reduction. A twin top mass $m_{\topt} \simeq m_t/\sqrt{\xi}$, where $\xi = v^2/f^2$, is predicted, while the resonances can be pushed to $m_* \sim 5 \TeV$, beyond LHC reach and without compromising the fine-tuning of the EW scale. Consequently, the collider phenomenology of \tH models is substantially different from that of a standard composite-NGB Higgs, see e.g.~\cite{Burdman:2014zta,Craig:2015pha,Curtin:2015bka,Barbieri:2016zxn,Chacko:2019jgi}.

In this paper, we present a new realization of the \tH paradigm, in fact the last missing construction with minimal NGB content and a bona-fide \tH mechanism, i.e.~with a twin parity that could be enforced exactly. The model is based on a global $SO(7)$ symmetry broken to the exceptional group $G_2$, and its main phenomenological characteristic is that the twin fields carry hypercharge, and in fact they have the same electric charge as the SM states.
This feature gives rise to new and exciting signals at colliders, which include long-lived charged twin scalars, twin quarks behaving as microscopic quirks, that after pair production form bound states that annihilate into SM final states, dileptons and $\gamma h$, $Z h$, or twin gluons, and some twin glueballs which, while likely stable on collider scales, eventually decay almost exclusively to diphotons. Albeit some of these signatures have been already experimentally explored, here they are linked to the naturalness of the EW scale, a strong motivation that warrants further exploration at the LHC, in particular in its high-luminosity phase: since production cross sections are of electromagnetic size, large data samples mean excellent prospects to probe this scenario.

The same structural reason that requires the SM particles and their twins to have identical electric charges, also explains the absence of a twin photon. Because the global symmetry of the composite sector is a rank three $SO(7)$, there is no place to embed both the SM and twin hypercharge $U(1)$, after the SM weak $SU(2)_L$ and its twin $SU(2)_{\Lt}$ have been accommodated; by construction the photon and its twin must be identified. The \tH model presented here is therefore smaller, in terms of its global symmetries, than previous constructions based on $SO(8)/SO(7)$ \cite{Geller:2014kta,Low:2015nqa,Barbieri:2015lqa}, while retaining custodial symmetry. There is however a tradeoff: if the $SU(2)_{\Lt}$ is gauged by twin weak gauge bosons, a $Z'$ coupled to the SM fermions is predicted, for which there exist strong experimental bounds from the LHC, rendering the model less plausible. This issue is resolved in the same way as in standard \tH constructions \cite{Craig:2015pha,Barbieri:2015lqa}, where the twin photon is avoided  by decoupling the twin gauge bosons at energies above the confinement scale. Such a departure from a complete, or fraternal \cite{Craig:2015pha}, implementation of twin parity has the advantage of introducing the required source of explicit $Z_2$ symmetry to misalign the vacuum with respect to the twin symmetric one, i.e.~$\xi \lesssim 1/2$, as Higgs couplings measurements and EW precision constraints require. This the minimal scenario is therefore preferred from both phenomenological and theoretical considerations; yet we find the study of the fraternal version of the model illustrative at the very least.

Finally, let us briefly comment on how our exceptional twin Higgs compares with previous models of neutral naturalness. Because the twin quarks are uncolored but carry hypercharge as well as twin color, their phenomenological signatures resemble those of folded supersymmetry \cite{Burdman:2006tz} or the quirky little Higgs \cite{Cai:2008au}. However, the scenario presented in this paper is not supersymmetric, and it enjoys a custodial symmetry, in contrast to the quirky little Higgs. Furthermore, because the twin sector is hypercharged but $SU(2)_L$-neutral, its phenomenology shows a novel predilection for photons and $Z$'s. Particularly, the twin top, which is the minimal light degree of freedom for a successful implementation of the \tH mechanism, can be produced at colliders with $\gamma, Z$-mediated cross sections that, although small, could be eventually probed by the HL-LHC, for instance in $\gamma h$, $Z h$ final states.

The rest of the paper is organized as follows. In the next section we present the symmetry structure and NGBs of the composite sector giving rise to the exceptional twin Higgs. In Section~\ref{weak} we show how the elementary gauge and fermion fields are coupled to the strong dynamics, both for the SM particles and their twins. We study two different options regarding the latter: in the first (fraternal), all the twin global symmetries are gauged, thus it includes, beyond a twin top, a twin bottom and twin leptons, while in the second (minimal), no twin gauge bosons are considered and the twin NGBs remain uneaten. The scalar potential, for the Higgs and for the extra scalars (in the minimal realization) is analyzed in Section~\ref{potential}. Lastly, the exceptional phenomenology of the model is presented in Section~\ref{pheno}, where both indirect and directs constraints are discussed, as well as the future prospects for discovery.


\section{Higgs sector} \label{strong}

The \tH mechanism relies on the Higgs arising as a NGB from the spontaneous breaking of a global symmetry.
In this work we assume that such a breaking is driven by a strongly interacting sector that confines at a scale $m_*$ close to the $\TeV$.
The characteristic mass and coupling of the composite resonances is set by $m_*$ and $g_*$ respectively, related by the symmetry-breaking order parameter, $f$, as $m_* = g_* f$.

The novel features of our scenario stem from a different global symmetry breaking pattern, $SO(7) \to G_2$, with respect to previous \tH models. 
This breaking has the peculiarity of being the smallest that gives rise to seven NGBs while leaving an unbroken custodial symmetry, $SU(2)_L \times SU(2)_R \subset G_2$.%
\footnote{The other relevant 7-spheres are $SO(8)/SO(7)$, $SU(4)/SU(3)$ and $SO(5)/SO(3)$, where only the first respects custodial symmetry. The coset $SO(7)/G_2$ has also been explored in \cite{Chala:2012af,Ballesteros:2017xeg} in the context of ordinary composite-Higgs models.}
The \tH mechanism can still be operative in this coset, with the Higgs complex doublet $H$ and its twin $\Ht$ embedded in the spinorial $\mathbf{8}$ representation of $SO(7)$.
Interestingly, $SO(7)$ has rank three and it contains as a subgroup the product $SU(2)_L \times SU(2)_{\Lt} \times SU(2)_{\Z}$, out of which the unbroken $SU(2)_R$ is the diagonal combination of the $SU(2)_{\Lt}$ and $SU(2)_{\Z}$.
As it will become clear in the following, a discrete $Z_2$ symmetry, a.k.a.~twin parity, requires the SM particles and their twins to share the same quantum numbers under $SU(2)_{\Z}$, which in turn implies they have identical electric charges, with the SM photon being its own twin.

We parametrize the $SO(7)/G_2$ coset with the $\Sigma$ vector
\beq
\Sigma = U(\pi) \Sigma_0 = \begin{pmatrix}
\pi_1 & \pi_2 & \pi_3 & \pi_4 & \pi_5 & \pi_6 & \pi_7 & \sigma
\end{pmatrix}^T \, , \,\,\,
\sigma = \sqrt{1-\pi_{\hat a}^2} \, ,
\label{sigma}
\eeq
associated with the vacuum expectation value of a spinor of $SO(7)$, that is $\Sigma \sim \mathbf{8}$.%
\footnote{$SO(7)$ matrices in the spinor representation can be found in App.~\ref{embed}. As usual the Goldstone matrix is given by $U(\hat \pi) = \exp ( i \sqrt{8/3} \, \hat \pi_{\hat a} T^{\hat a})$, and to arrive at \Eq{sigma} we redefined $\pi_{\hat a} = \hat \pi_{\hat a} \sin \widehat \Pi$, with $\widehat \Pi = \sqrt{\hat \pi_{\hat a} \hat \pi^{\hat a}}$, while $\Sigma_0 = (0 \,\, 0 \,\, 0 \,\, 0 \,\, 0 \,\, 0 \,\, 0 \,\, 1 )^T$.}
Given the decomposition $\mathbf{8} = (\mathbf{2},\mathbf{1},\mathbf{2}) + (\mathbf{1},\mathbf{2},\mathbf{2})$ under $SU(2)_L \times SU(2)_{\Lt} \times SU(2)_{\Z}$, the Higgs and its twin are identified as
\beq
\label{higgses}
(\mathbf{2},\mathbf{1},\mathbf{2}): \, H 
= {f \over \sqrt{2}} \begin{pmatrix} \pi_2 + i \pi_1 \\ h - i \pi_3 \end{pmatrix} \, , \quad
(\mathbf{1},\mathbf{2},\mathbf{2}): \, \Ht 
= {f \over \sqrt{2}} \begin{pmatrix} \pi_6 + i \pi_5 \\ \sigma - i \pi_7 \end{pmatrix} \, ,
\eeq
where $h \equiv \pi_4$. It then follows that under the custodial $SU(2)_L \times SU(2)_R$ symmetry, the Higgs has the proper quantum numbers, \ie $H \sim (\mathbf{2},\mathbf{2})$, while the twin Higgs decomposes as $\Ht \sim (\mathbf{1},\mathbf{1}) \oplus (\mathbf{1},\mathbf{3})$, which correspond to the singlet radial component, $\sigma$, and a triplet of $SU(2)_R$.
Their kinetic terms take the same form as in other spherical cosets,
\beq
\frac{f^2}{2} |\partial_\mu \Sigma|^2 = \frac{f^2}{2} (\partial_\mu \pi_{\hat a})^2 + \frac{f^2}{2} \frac{(\pi_{\hat a} \partial_\mu \pi_{\hat a})^2}{1-\pi_{\hat b}^2} \,.
\label{kinterm}
\eeq

Twin parity in the strong sector is realized as a discrete $SO(7)$ transformation of the form
\beq
\mathcal{P} = \begin{pmatrix}
 & \mathbb{1}_4  \\
\mathbb{1}_4 &  
\end{pmatrix} \,,
\label{parity}
\eeq
whose action on $\Sigma$ enforces $H \leftrightarrow \Ht$. This parity also interchanges the $SU(2)_L$ and $SU(2)_{\Lt}$, while leaving the $SU(2)_{\Z}$ intact. 
This explains why the twin NGBs, $\wt^\pm \equiv f (\pi_6 \pm i \pi_5)/\sqrt{2}$ and $\wt_0 \equiv f \pi_7$ have the same electric charge as those eventually eaten by the $W^\pm$ and the $Z$.

Let us finally recall that in composite-\tH models the strong sector is also symmetric under $SU(3)_C \times SU(3)_{\Ct} \times Z_2$ transformations, with the $Z_2$ exchanging $C \leftrightarrow \Ct$. Moreover, to reproduce the hypercharges of the SM fermions an extra $U(1)_X$ abelian symmetry is introduced, along with its counterpart $U(1)_{\Xt}$ and with twin parity enforcing $X \leftrightarrow \Xt$.%
\footnote{There is a more minimal alternative consisting in a single $U(1)_X$ shared by the SM and their twins, with the $Z_2$ acting trivially on it. We however opt for the introduction of $U(1)_{\Xt}$ in order to distinguish the composite operators that couple to the SM (right-handed) leptons from their twins, see below.}


\section{Gauge and fermion sectors} \label{weak}

To specify the couplings of the external (i.e.~weakly coupled in comparison to $g_*$, elementary) SM gauge and fermion fields and their twins to the strong sector, we need to identify the operator content of the latter. Besides the scalar operator responsible for the spontaneous breaking of $SO(7)$ to $G_2$, there is a vector current associated to each global symmetry of the strong dynamics: $\mathcal{J}^A \sim \mathbf{21}$ of $SO(7)$, which under $G_2$ decomposes in unbroken $\mathcal{J}^a \sim \mathbf{14}$ and broken $\mathcal{J}^{\hat a} \sim \mathbf{7}$ components, $\mathcal{J}_\mu^{X,\Xt}$, and the color and twin color currents $\mathcal{J}_\mu^{C,\Ct}$. The weak gauging of a given global symmetry then gives rise to the coupling of the elementary gauge fields with the corresponding composite currents, \ie $g_i A_\mu^i \mathcal{J}^{\mu i}$.
Besides, in order to implement the mechanism for fermion mass generation known as partial compositeness \cite{Kaplan:1991dc}, we assume the strong sector contains fermionic operators $\Psi$ in non-trivial representations of $SO(7) \times SU(3)_C \times SU(3)_{\Ct} \times U(1)_X \times U(1)_{\Xt}$. The interaction of the elementary fermion fields $\psi$ with the strong dynamics proceeds via linear couplings of the form $y \bar \psi \Psi$. The actual set of fermionic operators depends on which twin fermions are present at energies around $m_*$, that in turn depends on which of the global symmetries are gauged. We discuss the most interesting possibilities in the following. 


\subsection{Fraternal} \label{fraternal}

A fully fledged analog of the standard \tH models, which are based on a global $SO(8)$ symmetry and where twin partners are introduced for all the SM gauge bosons and for at least one complete generation of SM fermions \cite{Craig:2015pha}, can also be constructed upon the symmetry  group $SO(7)$.
In this case however, because of the smaller rank of $SO(7)$ compared to $SO(8)$, the gauge content of the twin sector is necessarily reduced. 
In particular, elementary massless vectors gauge $SU(2)_L \times SU(2)_{\Lt} \times U(1)_{\Y}$, with 
\beq
\Y = T_{\Z}^3 + X + \Xt \, ,
\label{hyperhat}
\eeq
where $T_{\Z}^3$ is the diagonal generator $U(1)_{\Z} \subset SU(2)_{\Z}$.%
\footnote{Should one give up on exact twin parity, different global symmetries could be gauged, e.g.~$\Y = T_{\Z}^3 + X$, in which case the twins would have different hypercharges than the ones considered in this work.}

\begin{figure}[!t]
\begin{center}
\includegraphics[width=3.5in]{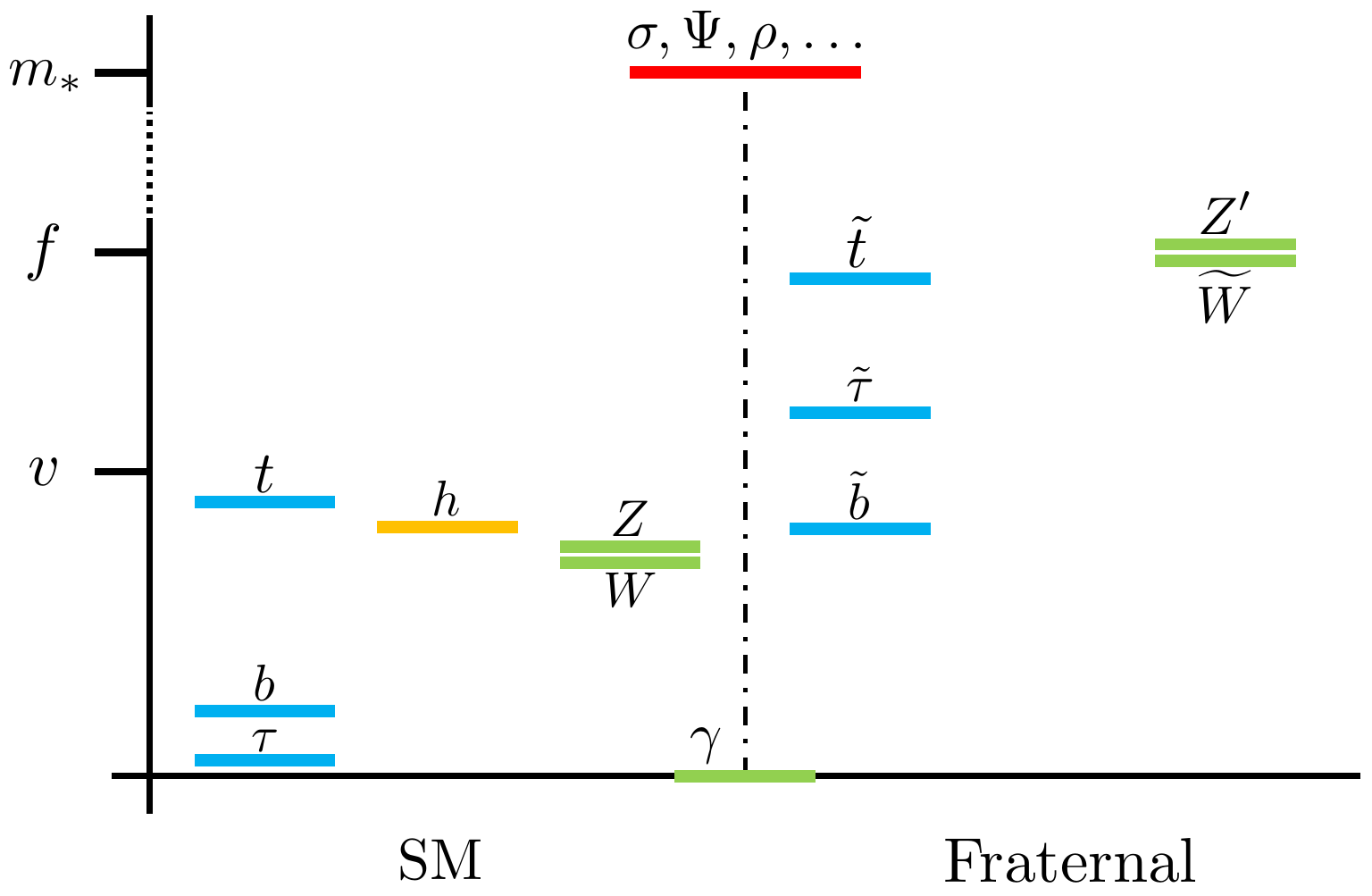}
\caption{Sketch of the fraternal spectrum of the exceptional twin Higgs.}
\label{fig:fra}
\end{center}
\end{figure}

Since $\vev{\Ht} \sim f$ spontaneously breaks $SU(2)_{\Lt} \times U(1)_{\Z}$ to $U(1)_R \subset SU(2)_R$, the three gauge bosons associated to the broken directions become massive, while the unbroken hypercharge group is identified with the diagonal combination of $U(1)_{\Lt} \times U(1)_{\Y}$, where $U(1)_{\Lt} \subset SU(2)_{\Lt}$, that is 
\beq
Y = T_R^3 + X + \Xt \, , \quad T_R^3 = T_{\Lt}^3 + T_{\Z}^3 \,.
\label{hyper}
\eeq
Only after the Higgs field develops a VEV, $\vev{H} \sim v$, does the $SU(2)_L \times U(1)_Y$ EW symmetry gets broken to $U(1)_Q$ and the $W$ and $Z$ become massive. From such a pattern of symmetry breaking it follows that electric charge is given by
\beq
Q = T_{L}^3 + Y = T_{L}^3 + T_{\Lt}^3 + T_{\Z}^3 + X + \Xt \, ,
\label{emcharge}
\eeq
thus twin parity implies that the SM particles and their twins have the same electric charge.

The masses of the gauge bosons follow from \Eq{kinterm} upon introducing covariant derivatives $\partial_\mu \to D_\mu = \partial_\mu - i g_i A_\mu^i T^i$, with $g_i A_\mu^i T^i = g W_\mu^\alpha T_L^\alpha + \gt \Wt_\mu^\alpha T_{\Lt}^\alpha + \gh \Ah_\mu \Y$. In the unitary gauge,
\bea
\frac{f^2}{2} |D_\mu \Sigma|^2 \!\!\!&=&\!\!\! \frac{f^2}{2} \frac{(\partial_\mu h)^2}{1-h^2} 
+ \frac{g^2 f^2}{4} h^2 \left[ W_\mu^+ W^{\mu -} + \left( \frac{\gh}{g} \Ah_\mu - W^3_\mu \right)^2 \right] \nn \\
&& \qquad \qquad + \, \frac{\gt^2 f^2}{4} (1 - h^2) \left[ \Wt_\mu^+ \Wt^{\mu -} + \left( \frac{\gh}{\gt} \Ah_\mu - \Wt^3_\mu \right)^2 \right] \,,
\label{gaugeterm}
\eea
where $h = (v + \hat{h})/f$, with $v \approx 246 \GeV$ and $\hat{h}$ the Higgs boson. This equation clearly shows the upshot of twin parity, which recall exchanges $SU(2)_L$ and $SU(2)_{\Lt}$ and $h \leftrightarrow \sigma = \sqrt{1-h^2}$. As a result, we can consistently impose a discrete $Z_2$ symmetry acting on the elementary gauge fields as $W_\mu^\alpha \leftrightarrow \Wt_\mu^\alpha$ and trivially on $\Ah_\mu$, which enforces $g = \gt$ while leaving $\gh$ free.
This constitutes a novel implementation of the \tH mechanism in the gauge sector, leading to important phenomenological departures from the SM, but no (massless) twin photon.
To see this right away, let us work in the limit $\xi \equiv v^2/f^2 \to 0$, that is neglecting EW symmetry-breaking (EWSB) effects. The SM hypercharge gauge field is given by $B = c_{\thetat} \Ah + s_{\thetat} \Wt^3$ with $\tan{\thetat} = s_{\thetat}/c_{\thetat} = \gh/\gt$ and gauge coupling $g' = \gt s_{\thetat}$. 
The orthogonal combination, $Z' = c_{\thetat} \Wt^3 - s_{\thetat} \Ah$, gets a mass term
\beq
m_{Z'} = \sqrt{\gt^2 + \gh^2} \frac{f}{2} = \frac{\gt}{c_{\thetat}} \frac{f}{2} 
\label{Zpmass}
\eeq
and, by virtue of the mixing $s_{\thetat}$ with $\Ah$, it couples to the SM fermions, which carry non-zero $\Ah$-charges (the same as their twins), given by $\Y = Y - T_{\Lt}^3 = Q - (T_L^3 + T_{\Lt}^3)$, see Table~\ref{table:qn}.
As we will show in Section~\ref{Zprime}, such a $Z'$ contributes to the EW precision tests (specifically to the $Y$-parameter) and, most importantly, can be produced at the LHC with significant cross sections.%
\footnote{Including EWSB terms does not change this conclusion, because the associated corrections to the mass and couplings of the $Z'$ are suppressed by $\xi \lesssim 0.1$.} The only other extra gauge boson, the $\Wt^{\pm}$, gets a mass $m_{\Wt} = \gt f \sqrt{1-\xi} /2$ and it has electric charge $\pm 1$, being the twin of the SM $W^\pm$.
The contribution of the twin gauge bosons to the Higgs potential will be discussed in Section~\ref{potential}. 

\begin{table}[t]
\begin{center}
\begin{tabular}{l | ccc | c}
& $SU(2)_L$ & $SU(2)_{\Lt}$ & $U(1)_{\Y}$ & $U(1)_Y$ \\
\hline
$H$ & $\mathbf{2}$ & $\mathbf{1}$ & $1/2$ & $1/2$ \\ 
$\Ht$ & $\mathbf{1}$ & $\mathbf{2}$ & 1/2 & $\pm 1\, , \, 0$ \\ 
\hline
$W$ & $\mathbf{3}$ & $\mathbf{1}$ & $0$ & $0$ \\
$\Wt$ & $\mathbf{1}$ & $\mathbf{3}$ & $0$ & $\pm 1\, , \, 0$ \\
$\Ah$ & $\mathbf{1}$ & $\mathbf{1}$ & $0$ & $0$ \\
\hline
$q_L$ & $\mathbf{2}$ & $\mathbf{1}$ & $1/6$ & $1/6$ \\ 
$t_R$ & $\mathbf{1}$ & $\mathbf{1}$ & $2/3$ & $2/3$ \\ 
$b_R$ & $\mathbf{1}$ & $\mathbf{1}$ & $-1/3$ & $-1/3$ \\ 
$\qt_L$ & $\mathbf{1}$ & $\mathbf{2}$ & $1/6$ & $\frac{2}{3} \, , \, -\frac{1}{3}$ \\ 
$\topt_R$ & $\mathbf{1}$ & $\mathbf{1}$ & $2/3$ & $2/3$ \\ 
$\tilde b_R$ & $\mathbf{1}$ & $\mathbf{1}$ & $-1/3$ & $-1/3$ \\ 
\hline
\hline
$\ell_L$ & $\mathbf{2}$ & $\mathbf{1}$ & $-1/2$ & $-1/2$ \\ 
$\tau_R$ & $\mathbf{1}$ & $\mathbf{1}$ & $-1$ & $-1$ \\ 
$\tilde \ell_L$ & $\mathbf{1}$ & $\mathbf{2}$ & $-1/2$ & $0 \, , \, -1$ \\ 
$\tilde \tau_R$ & $\mathbf{1}$ & $\mathbf{1}$ & $-1$ & $-1$
\end{tabular}
\end{center}
\caption{EW gauge quantum numbers of the SM fields and their twins. In the minimal scenario, with no twins for the SM $W$ and $Z$, only $SU(2)_L \times U(1)_Y$ are gauged and the twin low-energy content reduces to $\topt_{L,R}$ and $\wt^{\pm,0} \subset \Ht$.
\label{table:qn}}
\end{table}

Regarding the elementary fermions, anomaly cancellation requires a twin partner for every SM fermion, at least for one complete generation: in practice this means for the third generation, since the twin top is the crucial player leading to a successful cancellation of the Higgs potential induced by the top. 
The EW gauge quantum numbers of the twin particles simply follow from twin parity and are given in Table~\ref{table:qn}. Of particular importance for the phenomenology of this scenario is the fact that the twin fermions have the same electric charge as their SM partners (since twin fermions are $SU(2)_L$ neutral).

Besides, as in previous \tH models we assume the twin quarks are not colored but instead carry twin color, \ie they are fundamentals $\mathbf{3}$ of $SU(3)_{\Ct}$, which is gauged by the twin gluons. Twin parity exchanges them with the SM gluons, from where it follows $g_3 = \gt_3$, up to terms that explicit break the twin parity.%
\footnote{A natural alternative is that twin color gets spontaneously broken by the same strong dynamics at a scale close to $f$ \cite{Batell:2015aha}. We leave for a future work the study of such a scenario and its phenomenological consequences associated with massive twin gluons \cite{heavytgluon}.}

The SM fermions and their twins get mass via their interactions with the strong sector, that we assume to be of partial compositeness type, at least for the top sector. The top quark fields $t_R$, $q_L$ and their twins $\topt_R$, $\qt_L$ then linearly couple, with strengths $y_{L,R}$ and $\yt_{L,R}$, to composite operators. For a proper implementation of the \tH mechanism, the latter should transform under $SO(7) \times SU(3)_C \times SU(3)_{\Ct} \times U(1)_X \times U(1)_{\Xt}$ as $\Psi_t \sim (\mathbf{1},\mathbf{3},\mathbf{1})_{({2 \over 3},0)}$, $\Psi_q \sim (\mathbf{8},\mathbf{3},\mathbf{1})_{({2 \over 3},0)}$ and $\Psit_t \sim (\mathbf{1},\mathbf{1},\mathbf{3})_{(0,{2 \over 3})}$, $\Psit_q \sim (\mathbf{8},\mathbf{1},\mathbf{3})_{(0,{2 \over 3})}$. We recall that the spinorial representation of $SO(7)$ decomposes as $\mathbf{8} = \mathbf{1} + \mathbf{7}$ under $G_2$. 

At low energies the interactions of $q_L$ and $\qt_L$ can be simply obtained via the embeddings
\bea
Q_L \!\!\!&=&\!\!\! v_b b_L + v_t t_L = \frac{1}{\sqrt{2}} 
\begin{pmatrix}
i b_L & b_L & i t_L & - t_L & 0 & 0 & 0 & 0  
\end{pmatrix}^T \,, \nn \\
\Qt_L \!\!\!&=&\!\!\! \vt_b \bt_L + \vt_t \topt_L = \frac{1}{\sqrt{2}} 
\begin{pmatrix}
 0 & 0 & 0 & 0 & i \bt_L & \bt_L & i \topt_L & - \topt_L
\end{pmatrix}^T \,,
\label{topembed}
\eea
while those of $t_R$ and $\topt_R$ are trivial, being $SO(7)$ singlets. 
For instance, the top Yukawa couplings, generated at the scale $m_*$ with $y_t \sim y_L y_R/g_*$ and likewise for $\yt_t$, are given by
\beq
y_t f \bar t_R \Sigma^\dagger Q_L + \yt_t f \bar \topt_R \Sigma^\dagger \Qt_L + \hc = - y_t \bar t_R H q_L - \yt_t \bar \topt_R \Ht \qt_L + \hc \, ,
\label{yuk}
\eeq
from where the twin-top mass follows, $m_{\topt} = \yt_t f \sqrt{1-\xi}/\sqrt{2}$.
It is evident that a $Z_2$ symmetry acting on the elementary fields as $q_L, t_R \leftrightarrow \qt_L, \topt_R$ leads to $y_t = \yt_t$.
The rest of SM fermions and their twin partners can get masses in a similar manner, with twin parity enforcing the equality of their Yukawa couplings and thus $m_\psi = m_{\tilde \psi} \sqrt{\xi/(1-\xi)}$.
However, since as long as $\yt_\psi \ll y_t$ the associated contribution to the Higgs potential will be negligible (see Section~\ref{potential}), the approximate equality $y_\psi = \yt_\psi$ need not be enforced. 


\subsection{Minimal} \label{minimal}

The construction presented in the previous section demonstrates that it is feasible to build a \tH model based on $SO(7)$ as global symmetry and where twin parity could be exact. However, we advance here that the experimental constraints on the twin $Z'$ are quite severe, disfavouring such a twin-symmetric model in view of fine-tuning considerations. Besides, it is well known that the twin parity cannot be an exact symmetry at $m_*$ if the Higgs VEV is to be misaligned from the $Z_2$-symmetric vacuum $\vev{H}^2 + \vev{\Ht}^2 = f^2$. For these reasons we consider in this section a realization that is more minimal in terms of the (elementary) twin particle content.

We assume that the gauged $SU(2)_{\Lt}$ symmetry has been spontaneously broken at a scale $\Lambdat$ significantly above the strong sector confinement scale, such that at $m_*$ only $SU(2)_L \times U(1)_Y$ is gauged, in a similar fashion as the recent \tH constructions of \cite{Serra:2017poj,Csaki:2017jby}. This has several important consequences: the formerly eaten twin NGBs $\wt^{\pm, 0}$ are now physical and become massive (much in the same way as the QCD pions because of EWSB), and the twin fermionic content can be reduced to the essential, that is a (vector-like) twin top.

\begin{figure}[!t]
\begin{center}
\includegraphics[width=3.5in]{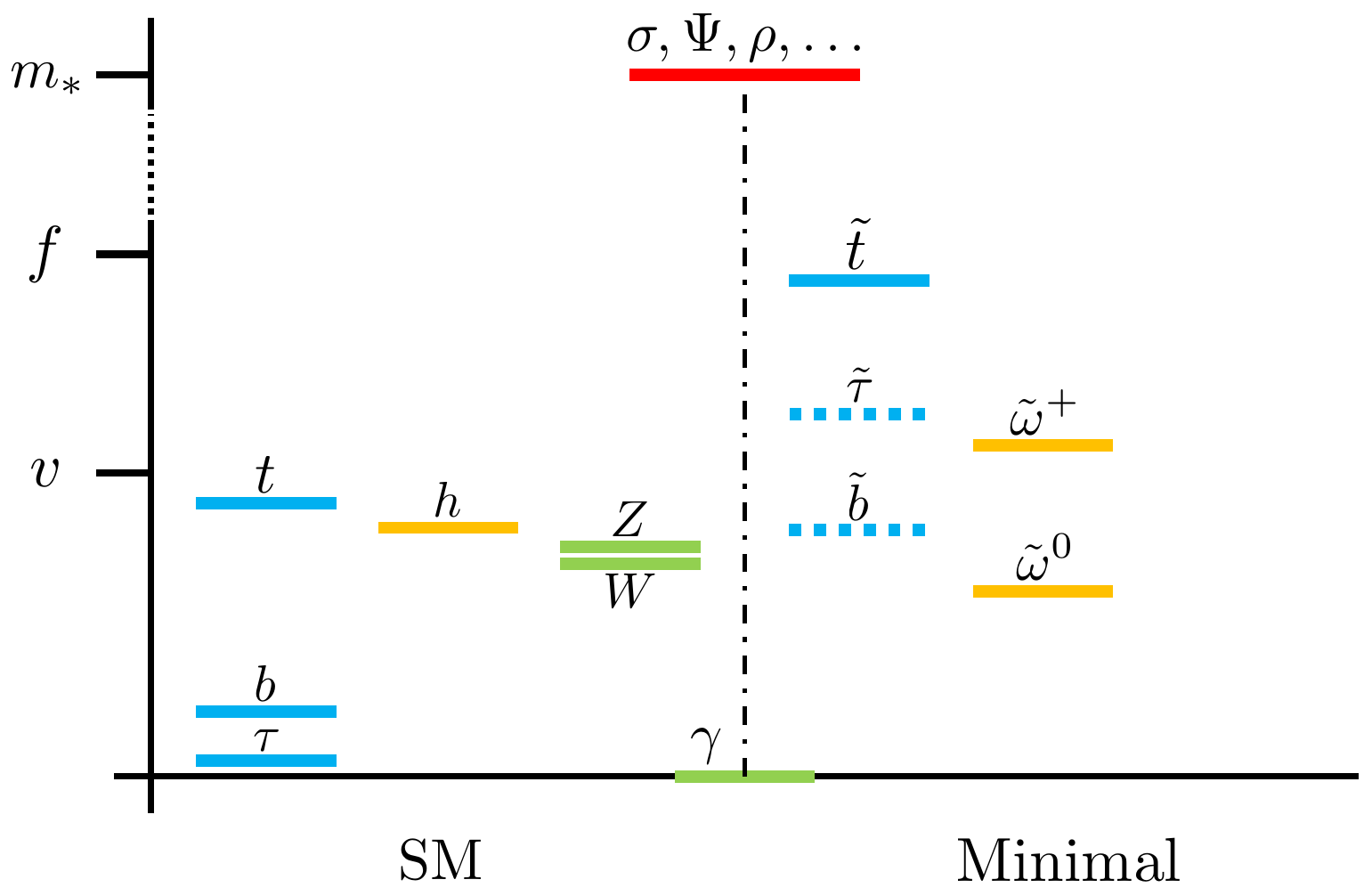}
\caption{Sketch of the minimal spectrum of the exceptional twin Higgs.}
\label{fig:fra}
\end{center}
\end{figure}

The kinetic terms for the twin Goldstones, along with derivative Higgs and self-interactions, follow from \Eq{kinterm},
\beq
\frac{f^2}{2} |D_\mu \Sigma|^2 = |D_\mu H|^2 + |D_\mu \wt^+|^2 + \frac{1}{2} (\partial_\mu \wt^0)^2 
+ \frac{\left( 2 \partial_\mu |H|^2 + 2 \partial_\mu |\wt^+|^2 + \partial_\mu (\wt^0)^2 \right)^2}{f^2 - 2 |H|^2 - 2 |\wt^+|^2 - (\wt^0)^2} \,,
\label{gaugeterm2}
\eeq
where $D_\mu \wt^+ = \partial_\mu \wt^+ - i g' B_\mu \wt^+$ ($\wt^+$ has electric charge $+ 1$, while $\wt^0$ is neutral). It is clear that in this scenario there can be no twin cancellation of the gauge radiative correction to the Higgs potential. However this can be kept small, \ie without compromise for fine-tuning, as long as EW composite resonances are sufficiently light (see Section~\ref{potential}).

The top and its twin couple to the strong sector as in \Eq{topembed}, where however we can consider the twin bottom is no longer present in the low-energy spectrum but decoupled above $m_*$,%
\footnote{Note that below $\Lambdat$, where the only gauged symmetries are the SM ones and twin color, the twin bottom as well as the twin top have vector-like charges.}
or that, even if it is light, its couplings differ from those of $\topt$. As a result the Yukawa couplings are now given by
\beq
- y_t \bar t_R H q_L - \frac{\yt_t}{\sqrt{2}} \bar \topt_R \left( i \wt^0 + \sqrt{f^2 - 2|H|^2 -2|\wt^+|^2 - (\wt^0)^2} \right)  \topt_L + \hc  \,.
\label{yuk2}
\eeq

Given the explicit breaking of the twin parity from the absence of the twin $\Wt$'s and the twin bottom (as well as the twins of the light flavors), it is no longer the case that $y_{L,R} = \yt_{L,R}$ (thus neither that $y_t = \yt_t$) at $m_*$. As we will show in Section~\ref{potential}, this indicates there is an upper bound on the scales (like $\Lambdat$) where the sources of $Z_2$ breaking originate, such that the twin cancellation of the top radiative correction to the Higgs potential is still effective. 


\section{Scalar potentials} \label{potential}

In this section we discuss the radiative generation of the Higgs potential and examine the conditions as well as the amount of fine-tuning required to achieve EWSB. We also compute the potential of the extra pseudo-NGBs in the case that no twin symmetries are gauged.\\

\textbf{Fraternal}: Let us discuss first the gauge contributions to the Higgs potential in the fraternal model, where the $SU(2)_L \times SU(2)_{\Lt} \times U(1)_{\Z}$ symmetries are gauged. The leading order (LO) terms in the gauge couplings arise at $O(g_i^2)$ from 1-loop diagrams with either the $W$'s, the twin $\Wt$'s or the $\Ah$, 
\beq
V_{g^2}^{\mathrm{UV}} = C_g \sum_i g_i^2 \, \Sigma^\dagger T^i T^i \Sigma = 
c_g \frac{3 m_\rho^2}{32 \pi^2} \frac{3}{2} \left( g^2 |H|^2 + \gt^2 |\Ht|^2 \right) \,,
\label{gaugepot}
\eeq
where we estimated the coefficient $C_g$ based on naive dimensional analysis (NDA), with an $O(1)$ uncertainty encoded in the parameter $c_g$.
This radiative correction is quadratically sensitive to the compositeness mass scale, which we parametrized by $m_\rho \sim g_\rho f/2 \lesssim m_*$. In the twin-symmetric limit $g = \gt$, \Eq{gaugepot} does not depend on the Higgs, which is the primary feature of \tH models. Note also that in our realization the twin cancellation of the $O(\gh^2)$ correction is automatic, since $\Ah$ is its own twin or, in other words, $H$ and $\Ht$ have identical $U(1)_{\Z}$ charges. In consequence, the usual hypercharge contribution to the Higgs potential is absent.

In contrast with most composite-NGB-Higgs models, (\ref{gaugepot}) is in fact the most relevant correction to the Higgs potential in our fraternal $SO(7)$ \tH model. This is because of the experimental constraints on the $Z'$, the mass eigenstate of the $\Wt^3$-$\Ah$ system up to $O(\xi)$ terms, see Section~\ref{Zprime} for the details. Since the couplings of the $Z'$ to SM fields scale with $s_{\thetat} = g'/\gt$ and its mass \Eq{Zpmass} is well approximated by $\gt f/2$, the bounds can only be satisfied either if the twin gauge coupling $\gt$ is substantially above $g$, thus breaking twin parity and spoiling the cancellation of \Eq{gaugepot}, or if $f$ is pushed to several TeVs. As we show in Section~\ref{Zprime}, either case implies significant fine-tuning of the Higgs potential.\\

\textbf{Minimal}: In view of these considerations, we now turn to the scalar potential in the minimal model, where only $SU(2)_L \times U(1)_Y$ are gauged. The LO gauge contribution formally reads as the l.h.s.~of \Eq{gaugepot}, but now evaluates to
\beq
V_{g^2}^{\mathrm{UV}} = c_g 
\frac{3 m_\rho^2}{64 \pi^2} \left[ (3 g^2 + g'^2) |H|^2 + 2 g'^2 |\wt^+|^2 \right] \,,
\label{gaugepot2}
\eeq
where the last term is from a $B$ loop, since the twin $\wt^\pm$ carries hypercharge. The contribution to the Higgs potential is similar to the LO gauge correction in standard composite-NGB-Higgs models, where in fact it is usually considered subleading and relatively unimportant compared to the corrections arising from the top (at least in partial compositeness). In contrast, \Eq{gaugepot2} is certainly important in the present \tH model, since it can introduce the required amount of $Z_2$-breaking to accomplish $v/f \ll 1$, as explained below. EW gauge corrections at $O(g_i^4)$ are generically subleading and we neglect them in the following.

The other important contributions to the scalar potential arise from the explicit $SO(7)$ symmetry-breaking terms sourced by the top and its twin, of which two different types are relevant: those generated at $m_*$, which we denote as UV (one such type, for gauge fields, is \Eq{gaugepot2}), and those generated from IR loops, independent of the details of the strong dynamics. Let us discuss them in turn.

The LO corrections from 1-loop UV diagrams with $q_L$ and $\topt_L$ arise at $O(y_L^2)$ and $O(\yt_L^2)$ and are given by~%
\footnote{There are no equivalent contributions at this order from $t_R$ or $\topt_R$ since neither $y_R$ nor $\yt_R$ break any shift symmetries, being both fields embedded in a singlet of $SO(7)$.}
\beq
V_{y^2}^{\mathrm{UV}} = C_y y_L^2 \sum_{\psi = t,b} \Sigma^\dagger v_\psi v_\psi^\dagger \Sigma + \widetilde C_y \yt_L^2 \Sigma^\dagger \vt_t \vt_t^\dagger \Sigma = 
c_y \frac{6 m_\Psi^2}{32 \pi^2} \left[ (y_L^2 - \yt_L^2) |H|^2 - \yt_L^2 |\wt^+|^2 \right] \,,
\label{topLpot}
\eeq
where we used NDA to estimate the size of $C_y = \widetilde C_y$, with the equality between the coefficients following from twin parity in the strong sector. Note that while the overall sign of this contribution (\ie the sign of $c_y = O(1)$) cannot be predicted without explicit information on the strong dynamics, a prediction is obtained instead for the relative sign of the mass terms of the Higgs and of the charged scalar. The latter arises because no $\bt_L$ loop has been included, since e.g.~it acquired a vector-like mass (\ie independent of $f$) $\mt_b > m_*$ \cite{Craig:2016kue} or because its couplings to the strong sector depart from those of $\topt_L$, these two states no longer related by a gauged $SU(2)_{\Lt}$ symmetry. In other words, if at $m_*$ a twin bottom were present (\ie in the fraternal model or if $\mt_b \ll m_*$) and its coupling to the strong sector were still fixed by $\yt_L$, then the $|\wt^+|^2$ term would be absent. In contrast, a potential for the neutral scalar $\wt^0$ is automatically absent because neither $y_L$ nor $\yt_L$ break the corresponding $U(1)_{\Lt-\Z}$ shift symmetry. Finally, an exact twin parity would enforce $\Delta y_L^2 \equiv y_L^2 - \yt_L^2 = 0$, thus exactly cancelling the $|H|^2$ term. This cancellation is the \emph{raison d'\^etre} of \tH models, and in the present realization does indeed take place at LO in the elementary (weak) couplings.
However, since the $Z_2$ symmetry is not exact, $\Delta y_L^2$ will generically be non-vanishing at the relevant scale, $m_*$, due to renormalization group evolution (RGE) from the $Z_2$-breaking couplings, \ie at NLO. In our case the latter are the EW gauge couplings $g, g'$, which contribute as
\beq
(\Delta y_L^2)_g = y_L^2 \frac{3 A g^2 + A' g'^2}{16 \pi^2} \log \frac{\Lambdat}{m_*} \,.
\label{dyLgauge}
\eeq
The coefficients $A$ and $A'$ parametrize our ignorance on the strong dynamics at scales above $m_*$ and are a priori $O(1)$ in size.
Another potential source of $Z_2$ breaking percolating to the top sector at one loop depends on the vector-like mass of the twin $b$. If $\mt_b \gg m_*$, then loops of $q_L = (t_L \,\, b_L)$ cannot be matched by those of $\topt_L$, inducing $(\Delta y_L^2)_b \sim (y_L^4 /16\pi^2) \log \mt_b/m_*$. We should note that this correction is model dependent, in particular there is no experimental reason to completely decouple the twin bottom. Besides, if twin colored fermions are decoupled, a differential running of the $SU(3)_{C}$ and $SU(3)_{\Ct}$ gauge couplings is induced, which eventually adds to $\Delta y_L^2$ (formally a 2-loop effect, \ie NNLO, but could easily be numerically important).

Other potentially relevant UV radiative corrections to the potential arise from 1-loop diagrams with 4 insertions of either $y_L$ or $\yt_L$,
\bea
V_{y^4}^{\mathrm{UV}} \!\!\!&=&\!\!\! D_y y_L^4 \Big( \sum_{\psi = t,b} \Sigma^\dagger v_\psi v_\psi^\dagger \Sigma \Big)^2 + \widetilde D_y \yt_L^4 \left( \Sigma^\dagger v_{\topt} v_{\topt}^\dagger \Sigma \right)^2 \nn \\
&=&\!\!\! d_y \frac{6}{32 \pi^2} \left[ y_L^4 |H|^4 + \yt_L^4 \left( f^2/2 - |H|^2 - |\wt^+|^2 \right)^2 \right] \,,
\label{topLpot2}
\eea
where $D_y = \widetilde D_y$ from twin parity. Even if $Z_2$ symmetric, (\ref{topLpot2}) contains both a Higgs mass and quartic terms. The latter could be particularly important in order to reproduce the physical Higgs mass, depending on the size of the IR contributions to the Higgs potential, which we discuss in the following.

Below $m_*$, loops of the top and its twin give further corrections to the scalar potential via the Yukawa couplings in (\ref{yuk2}). The leading logarithmic (LL) term reads as in standard \tH models,
\beq
V^{\mathrm{IR}}_{\mathrm{LL}} = \frac{3}{16 \pi^2} \left[ y_t^4 |H|^4 \log_t + \yt_t^4 \left( f^2/2 - |H|^2 - |\wt^+|^2 \right)^2 \log_{\topt} \right] \,,
\label{topIRpot}
\eeq
where $\log_t \equiv \log(m_*^2/m_t^2)$ and likewise for $\log_{\topt}$. This contribution to the Higgs potential is generically identified as the leading one in \tH models, being logarithmically enhanced in comparison to UV terms such as (\ref{topLpot2}). In our scenario the same would be true for the charged twin potential if it was not for \Eq{topLpot2}, which is quadratically sensitive to $m_*$ (there identified with $m_\Psi$).

Finally, we should note that NLL corrections to \Eq{topIRpot} from RGE due to the top/twin-top Yukawas and color/twin-color interactions have been shown in \cite{Greco:2016zaz,Contino:2017moj,Serra:2017poj} to be numerically important, in particular for extracting the physical Higgs mass. We expect a similar analysis could be performed in our scenario, leading to similar results, at least qualitatively if not quantitatively. In this work we simply bear in mind such corrections when presenting our $O(1)$ estimates in the next section.


\subsection{EWSB and Higgs mass} \label{ewsb}

The set of contributions to the Higgs potential presented above can be simply parametrized as (focusing only on the relevant component $h$) \cite{Barbieri:2015lqa},
\beq
V(h)/f^4 = \alpha h^2 + \beta \left( h^4 \log \frac{a}{h^2} + (1-h^2)^2 \log \frac{a}{1-h^2} \right)\, ,
\label{higgspot}
\eeq
where
\beq
\beta = \frac{3 y_t^4}{64 \pi^2} \, , \quad \log a = \log \frac{2 m_*^2}{y_t^2 f^2} + d_y \frac{y_L^4}{y_t^4}
\label{beta}
\eeq
and we have taken $\yt_t = y_t$ and $y_L = \yt_L$ in (\ref{topIRpot}) and (\ref{topLpot2}) respectively, which is a good approximation at the order we are working. The $Z_2$-breaking term $\alpha$ depends on whether the model is fraternal or minimal,
\bea
\label{alphaF}
\textrm{Fraternal:} \quad \alpha \!\!\!&=&\!\!\! c_g \frac{9 g_\rho^2 (g^2 - \gt^2)}{512 \pi^2} + c_y \frac{3 g_\Psi^2 \Delta y_L^2}{32 \pi^2} \,, \\
\textrm{Minimal:} \quad \alpha \!\!\!&=&\!\!\! c_g \frac{3 g_\rho^2 (3 g^2 + g'^2)}{512 \pi^2} + c_y A \frac{3 g_\Psi^2 y_L^2}{32\pi^2} \frac{3 g^2 + g'^2}{16 \pi^2} \log \frac{\Lambdat}{m_*} \,,
\label{alphaM}
\eea
where we have taken $m_\rho =g_\rho f/2$ and $m_\Psi = g_\Psi f$ in (\ref{gaugepot}, \ref{gaugepot2}) and (\ref{topLpot}), respectively, and assumed $A = A'$ in \Eq{dyLgauge}.

The first point to note is that in our scenario the Higgs quartic is approximately the same as in standard \tH models. In particular the IR contribution to the physical Higgs mass provides a significant fraction of the observed value,
\beq
\frac{(\delta m_h^2)^{\mathrm{IR}}}{m_h^2} = \frac{3 y_t^4 v^2}{8 \pi^2} \log \frac{m_*^4}{m_t^2 m_{\topt}^2} \approx 0.9 \, ,
\label{mh}
\eeq
where we have evaluated the top Yukawa at high scales, $y_t(1 \TeV) \approx 0.85$, in order to roughly include NLL effects, and taken $m_* = 5 \TeV$ and $m_{\topt}^2 \simeq m_t^2/\xi$ with $\xi = v^2/f^2 = 0.1$. It is then clear that UV corrections, proportional to $d_y$ in (\ref{beta}), can easily and naturally provide the missing fraction of $m_h$.

Therefore, to a good approximation the degree of fine-tuning required in our construction is determined by how unlikely it is to achieve a realistic EWSB, something that is directly controlled by the size of the $h^2$ terms in \Eq{higgspot}. As in most \tH models, there is a minimum amount of tuning ($\Delta$) as a direct consequence of twin parity: the $Z_2$-symmetric term in \Eq{higgspot} leads to $\Delta_{\mathrm{min}} = 2 \xi$, while $\xi$ is bounded from above from direct and indirect measurements of the Higgs couplings.
Another way to see this is that twin parity implies that the minimum of the potential is at $\vev{h^2} = \xi = 1/2$, thus some other finely-tuned $Z_2$-breaking contribution is needed to misalign the vacuum at $\xi \ll 1$. 
This is in fact the reason why sources of explicit $Z_2$-breaking are needed in \tH models, whose leading effect in the Higgs potential we have encoded in the term proportional to $\alpha$ ($\alpha > 0$ in order to accomplish such a misalignment). 
Note that in the case that $\alpha$ is controlled by a single $Z_2$-breaking source of the right size to reproduce a given value of $\xi$, then the associated tuning is simply given by $\Delta_{\mathrm{min}}$. However, in more complicated situations with several $Z_2$-breaking sources, it is possible that a tuning between them is needed to achieve a given $\xi$, thus increasing the overall tuning. A better measure of fine-tuning, applicable to either case, is $\Delta_i = m_h^2/4 \alpha_i f^2$.

\textbf{Fraternal}: The fraternal scenario illustrates both of these possibilities. The twin gauge coupling $\gt$ could be larger than $g$ such that, provided $c_g < 0$, the first term in (\ref{alphaF}) can be tuned to the $Z_2$-symmetric piece ($2 \beta \log a$), how much tuned determined by $\Delta_{\mathrm{min}}$; the hierarchy between $f$ and $v$ is however no longer determined by the bounds on the Higgs couplings, but by the bounds on the $Z'$ mass (Section~\ref{Zprime}), leading to $\Delta_{\mathrm{min}} \lesssim 1\%$ for $g_\rho \gtrsim 4$ (thus $m_\rho \gtrsim 6 \TeV$).
Otherwise, $f$ can be kept relatively low consistently with the $Z'$ bounds if there is another source of $Z_2$-breaking, e.g.~the second term in (\ref{alphaF}), and $\gt$ is sufficiently large.
In such a case the tuning is no longer tied to $v/f$, but instead is well approximated by $m_h^2/4 \alpha f^2$, clearly worse than $\Delta_{\mathrm{min}}$ for fixed $m_\rho$.

\textbf{Minimal}: The situation in the minimal model is certainly better, since
\bea
\label{tuneg}
\Delta_g \!\!\!&=&\!\!\! \frac{32 \pi^2}{3 c_g (3 g^2 + g'^2)} \frac{m_h^2}{m_\rho^2} \approx 15\% \frac{1}{c_g} \left(\frac{3 \TeV}{m_\rho}\right)^2 \,, \\
\Delta_y \!\!\!&=&\!\!\! \frac{128 \pi^4}{3 c_y A (3 g^2 + g'^2) y_L^2 \log_{\slashed{Z_2}}} \frac{m_h^2}{m_\Psi^2} \approx 30\% \frac{1}{c_y A} \frac{1}{y_L^2} \frac{10}{\log_{\slashed{Z_2}}} \left(\frac{4 \TeV}{m_\Psi}\right)^2 \,,
\label{tuney}
\eea
that is, none of the $Z_2$-breaking terms in (\ref{alphaM}) leads to a fine-tuning significantly worse than $\Delta_{\mathrm{min}} = 20\%$ for reasonable parameters: relatively heavy vector and fermionic resonances and a large separation between the $Z_2$-breaking UV scale and $m_*$, $\log_{\slashed{Z_2}} \equiv \log (\Lambdat/m_*)$ with e.g.~$\Lambdat/m_* = 2.5 \times 10^4$.


\subsection{Twin pseudo-NGBs} \label{tpNGBs}

Let us focus first on the mass of the charged twin scalar present in the minimal model. This arises from both gauge and top/twin-top radiative corrections, \Eq{gaugepot2} and \Eqss{topLpot}{topLpot2}{topIRpot} respectively. The largest of these is the UV $O(\yt_L^2)$ contribution in (\ref{topLpot}), which is quadratically sensitive to the cutoff, there parametrized by $m_\Psi$. As long as the coefficient $c_y$ is negative,%
\footnote{A coefficient $c_y < 0$ implies that the contribution from \Eq{topLpot} to $\alpha$ in the Higgs potential is also negative, assuming the perturbative result that the parameter $A$ in the RGE of $\Delta y_L^2$, \Eq{dyLgauge}, is positive. Even if this it at odds with the requirement (to misalign $\xi \ll 1$) that $\alpha > 0$, this is in fact not an issue, since $\alpha$ in the minimal scenario is easily dominated by the gauge term, as shown in \Eq{tuneg}.}
then the mass-squared is positive and of size
\beq
m_{\wt^+}^2 \sim \frac{3 m_\Psi^2 \yt_L^2}{16 \pi^2} \approx (550 \GeV)^2 \left(\frac{\yt_L}{1} \right)^2 \left(\frac{m_\Psi}{4 \TeV} \right)^2 \, .
\label{ctmass}
\eeq
The other corrections give rise to smaller contributions: the gauge contribution is suppressed by $O(g'^2/y_L^2)$ and yields $\Delta m_{\wt^+}^2 \sim (100 \GeV)^2 (m_\rho/3 \TeV)^2$, while the IR twin-top contribution (which is negative) is suppressed by $O(m_{\topt}^2/m_{\Psi}^2)$. After EWSB there also small corrections from the quartic coupling $|H|^2 |\wt^+|^2$, suppressed by $\xi$. Other contributions to the $\wt^+$ mass could also be present, e.g.~from loops of twin taus, if these are in the IR spectrum.

The attentive reader will have already noticed that none of the radiative corrections discussed above gave rise to a potential for the neutral twin scalar. This was to be expected, since neither gauging only $U(1)_Y$ nor considering a left-handed twin bottom with couplings different than those of the twin top (or just a decoupled $\bt$), breaks the $U(1)_{\Lt-\Z}$ shift symmetry protecting the $\wt^0$. However, an exact global $U(1)$ is not at all guaranteed, on the contrary quite generically a source of explicit symmetry breaking will be present, lifting the $\wt^0$. In fact, it is easy to imagine examples for such a source. One instance, perhaps not the best but certainly simple, is to consider a $f$-independent mass for the twin top $\mt_t \bar \topt_R \topt_L$, generated below $\Lambdat$ where no gauge symmetry forbids it, but naturally small $\mt_t \ll m_*$ since still protected by a chiral symmetry \cite{Craig:2016kue}. Such a mass in fact breaks explicitly $U(1)_{\Lt-\Z}$, and gives rise to an extra radiative correction to the scalar potential
\bea
V_{\mt}^{\mathrm{UV}} \!\!\!&=&\!\!\! \widetilde C_{\mt} \yt_L \yt_R \Sigma^\dagger \vt_t \mt_t + \hc \\ 
&=&\!\!\! c_{\mt} \frac{3 m_*^2}{16 \pi^2} \frac{y_t \mt_t \sqrt{2}}{f} \left[ \sin\!\gamma \, \wt^0 - \cos\!\gamma \sqrt{f^2 - 2|H|^2 -2|\wt^+|^2 - (\wt^0)^2} \right] \,, \nn
\label{vectorpot}
\eea
where we used NDA to estimate the size of $C_{\mt}$, $\yt_t \sim \yt_L \yt_R/g_* \simeq y_t$, and included a possible phase difference $\gamma$ between $\yt_t$ and $\mt_t$. Besides the tadpole term (which vanishes for $\sin \gamma = 0$), (\ref{vectorpot}) contains a $\wt^0$ mass of size (taking $\cos \gamma = 1$ and $c_{\mt} > 0$)
\beq
m_{\wt^0}^2 \sim (85 \GeV)^2 \left(\frac{\mt_t}{10 \GeV}\right) \left(\frac{750 \GeV}{f}\right) \left(\frac{m_*}{5 \TeV} \right)^2 \, .
\label{ntmass}
\eeq
The $f$-independent mass for the twin top is just one example of how the singlet twin could be lifted. Yet one should be aware that regardless of how $\wt^0$ gets a potential, it should be such that the Higgs potential is not significantly altered, not to raise the fine-tuning of the EW scale. In our example this would happen if $\mt_t \gtrsim 100 \GeV$, which explains why the reference value used in \Eq{ntmass}. Of course $\mt_t$ could be much smaller and the neutral twin much lighter accordingly. Let us note in this regards that $\wt^0$ does not linearly couple to SM fermions, but if kinematically allowed it opens a new Higgs-decay channel, which would force $f \gtrsim 1.2 \TeV$ to ensure consistency with LHC data.


\section{Phenomenology} \label{pheno}

In this section we discuss the collider signals of the exceptional \tH model. While we focus mostly on its minimal incarnation, the main constraints on the fraternal scenario will also be discussed at the end of this section to understand why such a case is less attractive. 

While sharing some of standard signatures of \tH models, such as the modification of the Higgs couplings and non-standard Higgs decays, these models present novel phenomenological features stemming from the fact that the twin particles carry hypercharge. We will first examine the indirect effects of the twin states on the well-measured properties of the SM particles. These are primarily the $Y$-parameter (which encodes new physics contributions to the hypercharge propagator), the running of the hypercharge gauge coupling $g'$, and the Higgs decay rate to photons. These departures from the SM arise at the 1-loop level (with the exception of those mediated by the twin $Z'$ in the fraternal model). Other indirect effects, common in constructions where the Higgs is a (custodially protected) composite pNGB, such as a the $S$-parameter or deviations in other Higgs couplings, will not be discussed here further and we will merely recall when necessary their implications on the parameters of the strong sector, $f$ and $m_* \sim m_\rho, m_\Psi$ \cite{Contino:2017moj}.%
\footnote{We will refrain as well from discussing the flavor aspects of our construction, which are in any case not significantly different than in other composite-\tH models \cite{Csaki:2015gfd}; see also \cite{KerenZur:2012fr,Frigerio:2018uwx} for a general discussion of flavor in partial compositeness.}

Direct production of the twin particles at colliders gives rise to the most interesting, though in some cases challenging, signatures of the exceptional twin Higgs. The twin quarks, being hypercharged, can be pair-produced via Drell-Yan (DY) and, because of the existence of twin strong interactions, exhibit the typical characteristics of the so-called quirks \cite{Kang:2008ea,Cai:2008au,Burdman:2008ek}. Such quirks are microscopic and lead to the formation of bound states that predominantly decay either to twin glueballs or back to the SM, while the glueballs in turn decay via an off-shell Higgs or, whenever this channel is forbidden, to photons.
Another attractive signature arises from the charged twin pNGB $\wt^\pm$ or the twin leptons: 
they are pair-produced in DY and likely stable on detector scales, thus subject to current long-lived charged particle (LLCP) searches.
Finally, we also present the stringent constraints on the twin $Z'$. These are the main reason why the minimal model is preferred, since such bounds are absent and one is left with a light but almost inert $\wt^0$, whose phenomenology is that of a twin axion-like particle that only couples to SM hypercharge, the study of which we defer to a future study.%
\footnote{As a matter of fact, $\omega_0$ would be a bona-fide twin axion if other sources of mass beyond the twin-color anomaly, such as that in \Eq{ntmass}, vanished. It would then solve the twin-strong-CP problem, if there was any to begin with.}


\subsection{Indirect effects} \label{indirect}

Since the twin particles carry hypercharge, they induce at the loop level a non-standard self-energy for the hypercharge field.%
\footnote{For the present purpose this is identified with the vector state coupled to the SM fermions via the usual $U(1)_Y$ current, that is the $\Ah$ field.}
At leading order in a momentum expansion, such type of corrections are customarily parametrized by the dimension-6 operator $-(Y/4m_W^2) (\partial_\rho B_{\mu \nu})^2$. Indeed, $N_{\psit}$ heavy fermions of mass $m_{\psit}$ and hypercharge $q_{\psit}$ generate a $Y$-parameter of size
\beq
Y_{\psit} = \frac{g'^2}{80\pi^2} \frac{m_W^2}{m_{\psit}^2} \Delta b_Y \, , \quad \Delta b_Y = \frac{4}{3} N_{\psit} q_{\psit}^2 \, .
\label{Ypsit}
\eeq
For e.g.~the twin top, $N_{\topt} = 3$, $q_{\topt} = \frac{2}{3}$ and $m_{\topt} \simeq m_t / \sqrt{\xi}$, leading to a very small contribution, $Y_{\topt} \sim 1 \times 10^{-5}$ for $\xi = 1/4$, in comparison to the per mille constraints from LEP \cite{Barbieri:2004qk} and more recently from the LHC \cite{Domenech:2012ai,Farina:2016rws}.
In fact, for any type of twin fermion we find that $Y_{\psit}$ is below LEP sensitivity for masses $m_{\psit} \gtrsim 100 \GeV$.
For an even lighter twin fermion, encoding its effects in the $Y$-parameter is no longer adequate, nor it is for the twin top at the LHC \cite{Farina:2016rws}. 
In these cases however, constraints could a priori still be placed by considering the contribution to the running of the hypercharge gauge coupling, encoded in the beta function coefficient $\Delta b_Y$ in (\ref{Ypsit}).
However, the present accuracy on neutral DY processes is too low to see any such effect \cite{Alves:2014cda}. 
The charged pNGB $\wt^\pm$ also gives a small contribution to the $Y$-parameter at one loop, which reads as in \Eq{Ypsit} with $m_{\psit}^2 \to 2 m_{\wt^+}^2$, see \Eq{ctmass}, and $\Delta b_Y = 1/3$.

More important are the 1-loop contributions of the twins to the Higgs decay to diphotons. In composite-Higgs models deviations of such a rate are generically $O(\xi)$ and a consequence of the modified couplings of the Higgs to the top and the $W$. The latter are also present in \tH models, since they are intrinsic of the NGB nature of the Higgs \cite{Giudice:2007fh}. On top of these, because the twin particles are electrically charged in our scenario, a direct $O(\xi)$ contribution is generated as well, specifically from a twin top loop. Since $m_{\topt} \gg m_h$, such an effect can be parametrized by the dimension-6 operator $(c_\gamma g'^2/m_W^2) |H|^2 B_{\mu \nu} B^{\mu \nu}$ with
\beq
(c_\gamma)_{\topt} = - \frac{q_{\topt}^2 g^2}{32 \pi^2} \frac{\xi}{1-\xi} \, .
\label{hggtwin}
\eeq
Note that in contrast to standard composite-Higgs models \cite{Giudice:2007fh}, this operator is not suppressed by $y_t^2/g_*^2$.%
\footnote{We have explicitly checked that such a suppression is found instead in the contribution to $c_\gamma$ from a loop of the charged pNGB $\wt^\pm$.}
The associated contribution to the $h\gamma\gamma$ vertex is $(c_{\gamma \gamma})_{\topt}/(c_{\gamma \gamma})_t \simeq -  \xi/(1-\xi)$, where $(c_{\gamma \gamma})_t$ is the top loop contribution in the SM; a result that matches the expectation from twin parity once $\xi = 1/2$. The contribution from the twin top therefore, being opposite in sign to the top's, increases by $O(\xi)$ the Higgs coupling to photons. Since the standard corrections lead to a reduction of $h \gamma \gamma$, $c_{\gamma \gamma} \simeq (c_{\gamma \gamma})_{\mathrm{SM}} \sqrt{1-\xi}$, the twin-top loop alleviates, rather than aggravates, the departure from the SM prediction, a fact that can become relevant as the precision on Higgs couplings measurements improves. 
Besides, since the twin top is uncolored, it does not affect the Higgs coupling to gluons, thus breaking the correlation between $h\gamma \gamma$ and $h gg$ usually found in composite-Higgs models.


\subsection{Direct production} \label{direct}

All our production cross sections have been computed using the MSTW2008NNLO parton distribution functions \cite{Martin:2009iq}.

\subsubsection{Long-lived charged particles} \label{llcp}

Some of the hypercharged twin particles are potentially stable on detector scales. This is the case of the charged twin scalar in the minimal model, since its leading interactions with the SM respect an accidental $Z_2$ symmetry $\wt^\pm \to - \wt^\pm$, while we expect the decay channels $\wt \to \topt + \bt , \taut + \tilde \nu_{\tau}$ to be kinematically closed, the latter because the experimental bound on a collider-stable twin tau is stronger than for $\wt$, as we show in the following (thus the decay $\taut \to \wt + \tilde \nu_{\tau}$ is open). Besides, we recall that minimality does not require neither the twin bottom nor the twin tau to be present in the IR, while in the fraternal model both of them are required for anomaly cancellation.

\begin{figure}[!t]
\begin{center}
\includegraphics[width=3.2in]{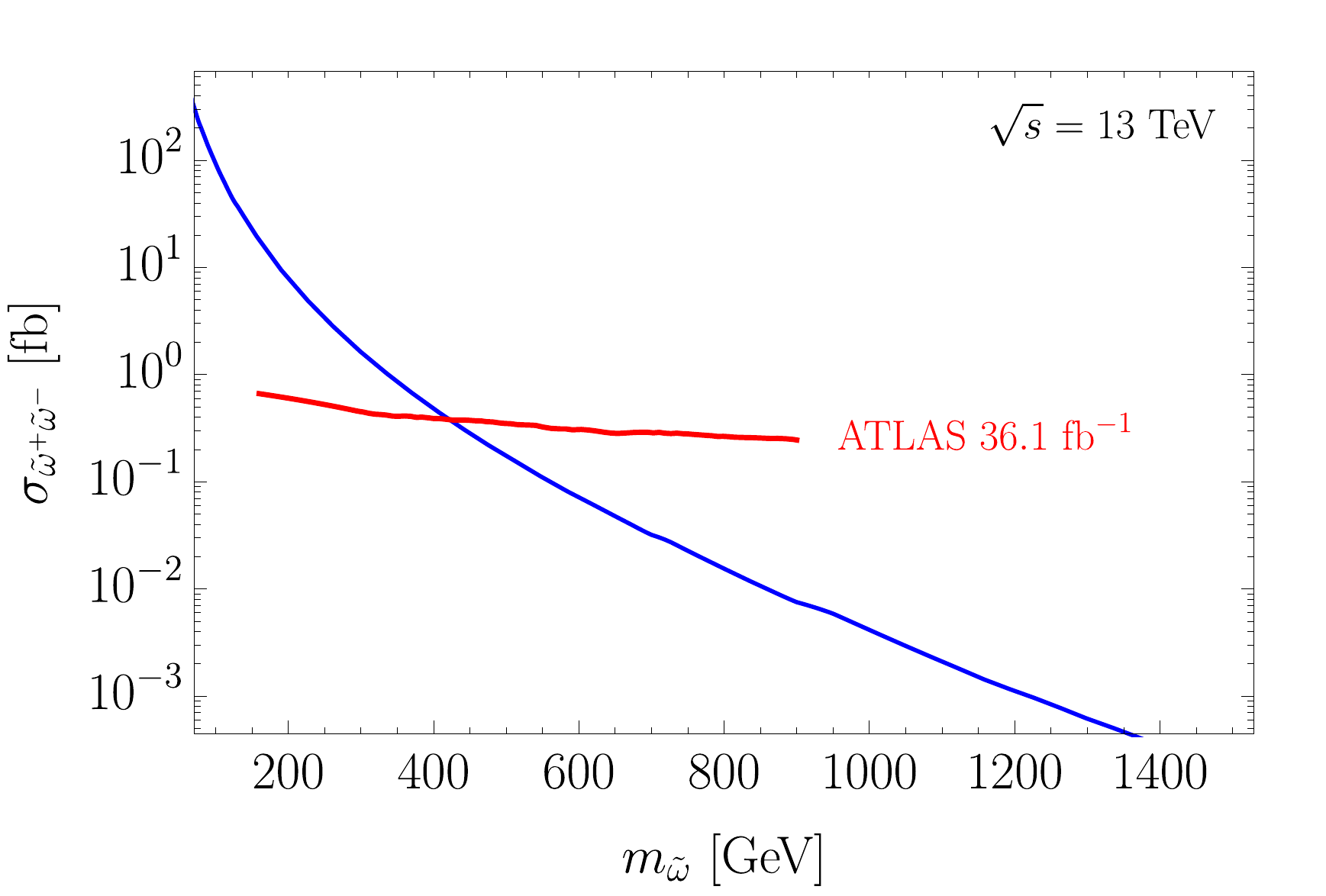}
\hspace{-1mm}
\includegraphics[width=3.2in]{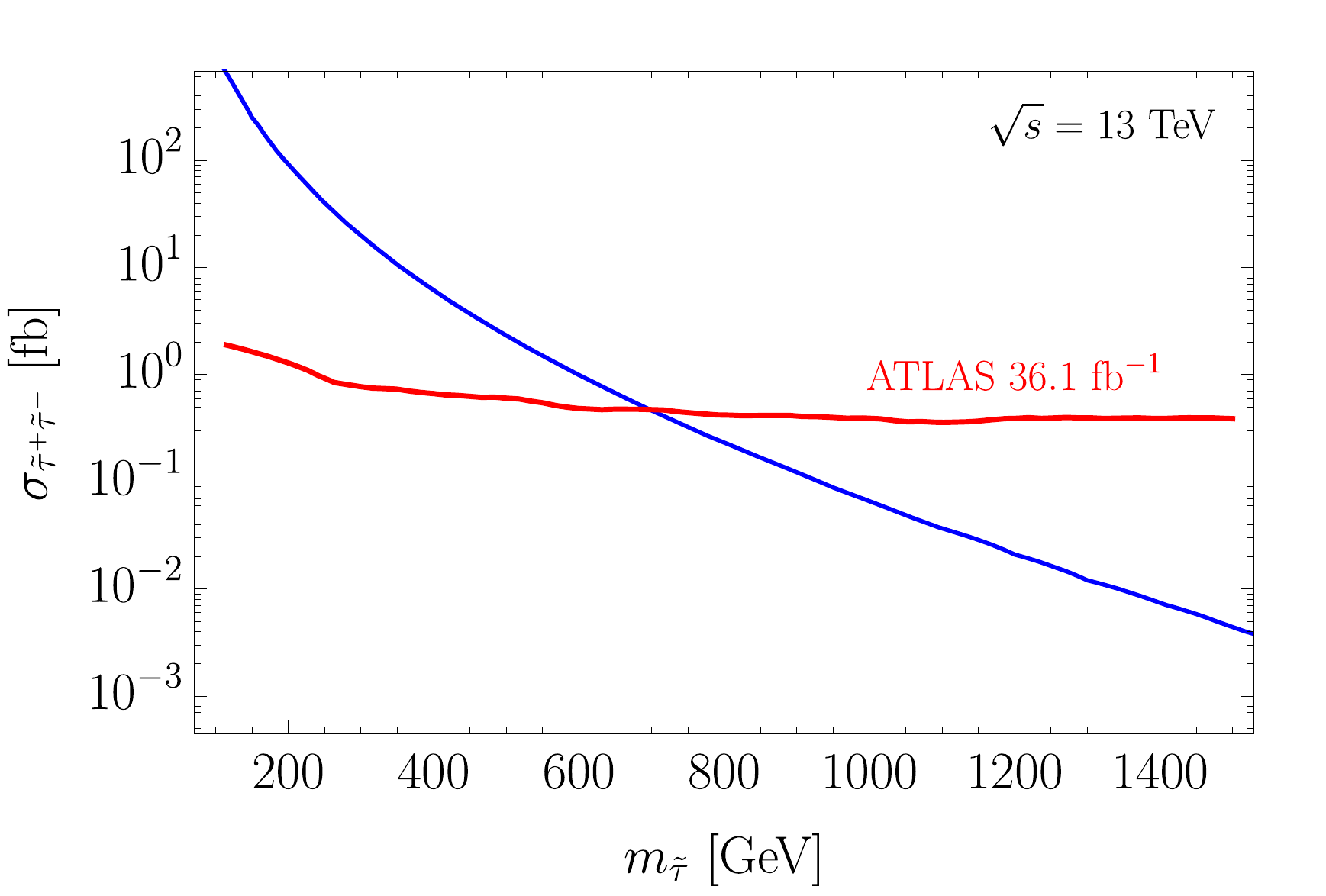}
\caption{Upper $95 \%$ CL limits on the cross section for DY pair-production of $|Q| = 1$ heavy stable scalars (left) and fermions (right) from an ATLAS search in  $36.1 \fb^{-1}$ of data at $\sqrt{s}= 13 \TeV$ \cite{Aaboud:2019trc} (red). The theoretical predictions for the twin pNGB $\wt^\pm$ and the twin tau $\taut$ are also shown (blue).}
\label{fig:llcp}
\end{center}
\end{figure}

Searches at the LHC for such type of particles have been carried out at $\sqrt{s} = 13 \TeV$, the latest analysis from ATLAS with $36.1 \fb^{-1}$ of data \cite{Aaboud:2019trc}.
In \Fig{fig:llcp} we show the corresponding constraints on the production cross sections for a $|Q| = 1$ charged scalar (left panel), identified with a stau in the ATLAS analysis, and a $|Q| = 1$ DY charged fermion (right panel), identified with a chargino. 
These constraints readily apply to our twin $\wt^+$ and $\taut$, respectively.
The comparison of the predicted cross section with the experimental bound for the former shows that masses below $420 \GeV$ are excluded. This should be compared with our expectation for the twin scalar mass $m_{\wt^+} \sim 550 \GeV$, \Eq{ctmass}. Even though the current bound does not quite reach such values, a factor of a few improvement in experimental sensitivity would start probing the relevant parameter space. 
For what regards the twin tau, the situation is certainly different, since twin parity would imply $m_{\taut} \simeq m_{\tau}/\sqrt{\xi}$, while data indicates that only masses above $700 \GeV$ are allowed. Of course, explicitly breaking twin parity in the lepton sector is a priori an option. However, the experimental bound is so stringent that the twin-tau Yukawa should be larger than the twin-top's to avoid it, thus a large contribution to the Higgs potential should be expected. This extra complication is yet another reason why the fraternal implementation of our scenario is disfavoured versus the minimal model, where the bounds on the charged scalar are less consequential and the $\taut$ could be lifted by giving it a vector-like mass.

While LLCPs could be regarded as a differential phenomenological feature of the exceptional twin Higgs, it is important to note that neither the stability of $\taut$ nor of $\wt^+$ are due to super-selection rules. Indeed, at low energies these states only carry electric charge and therefore decays such as $\taut \to \tau + \gamma, Z$ or $\wt \to \ell + \nu, q' + q, W + \gamma, Z$ are allowed. Let us discuss in some detail the decay of the twin tau (a similar discussion holds for $\wt^+$). The first point to note is that, since $\taut$ and $\tau$ have different $X$ (and $\Xt$) charges, the interaction mediating the decay cannot be generated by the strong dynamics alone, thus it should have its origin at some UV scale $\Lambda$. One instance is an interaction of the form $\bar \tau_R \sigma^{\mu \nu} \Ht \tilde \ell_L \Ah_{\mu \nu}$, 
which could be generated at low energies with a coefficient $\gh/m_* \Lambda$, leading to a $\taut$ decay rate
\beq
\Gamma_{{\taut} \to \tau + \gamma} \sim \frac{e^2 m_{\taut}^3}{8 \pi g_*^2 \Lambda^2} \approx (0.2 \, \mathrm{cm})^{-1} \left( \frac{m_{\taut}}{100 \GeV} \right)^3 \left( \frac{2 \pi}{g_*} \right)^2 \left( \frac{10^7 \GeV}{\Lambda} \right)^2 \, .
\label{tautdecay}
\eeq 
Therefore, model-dependent UV considerations could render the twin tau (and/or $\wt^+$) short-lived, displaced, or long-lived. It would be interesting to study these types of signatures at the LHC, e.g.~similar to those of an excited tau if the twin tau decays promptly. Finally, we restate that given the constraints in \Fig{fig:llcp}, the decay $\taut \to \wt + \tilde \nu_{\tau}$ would also proceed in the minimal model, and dominate over (\ref{tautdecay}).

\subsubsection{Hypercharged quirks} \label{quirks}

The lightest twin quark of the exceptional twin Higgs behaves as a quirk \cite{Kang:2008ea}, a heavy stable particle that interacts via a new unbroken non-abelian gauge group and carries SM charges, in our case twin color and hypercharge, respectively. 
Once pair-produced, via DY in our scenario, quirks do not hadronize but instead form (meta-)stable strings. This is because any of the twin quarks are heavy in comparison with the scale where twin-color interactions become strong, \ie $m_{\qt} \gg \Lqcdt$: in such a case the breaking of the string by pair production takes an exponentially large time, $t_{\textrm{break}} \sim (4 \pi^3/m_{\qt}) \exp(c \, m_{\qt}/\Lqcdt)^{2}$, with $c$ an $O(1)$ factor that depends on the precise definition of $\Lqcdt$. 
The absence of twin quarks lighter than $\Lqcdt$ follows from several considerations. On the one hand, naturalness of the Higgs potential indicates that the SM and twin-color gauge couplings are not substantially different at $m_*$, $\gt_s \approx g_s$, thus $\Lqcdt$ will not be far apart from the scale $\Lqcd \approx 250 \MeV$ where the SM color becomes strong. The difference arises mainly from $Z_2$-breaking radiative effects that originate from the different colored content of $SU(3)_C$ and $SU(3)_{\Ct}$ below $m_*$, since we assumed a single generation of twin quarks (and $m_{\qt} \neq m_{q}$). Such an effect leads to $\Lqcdt \sim 2 - 10 \GeV$, depending on $m_*$, on the exact values of $\gt_s$ and $g_s$ at $m_*$, and on the twin quark masses \cite{Craig:2015pha}. On the other hand, experimental constraints on the twin quarks force them to be considerably heavier than $\Lqcdt$. In particular, the precise measurement at LEP1 of the $Z$ decay width, $\Delta \Gamma_Z /\Gamma_Z \approx 9 \cdot 10^{-4}$ \cite{ALEPH:2005ab}, rules out twin quarks (with $Y_{\qt} = 2/3$ or $-1/3$) for which the decay channel $Z \to \bar{\qt} \qt$ is kinematically open, that is $m_{\qt} \lesssim m_Z/2 \approx 45 \GeV$. This constraint is only relevant for the twin bottom, since it implies $m_{\bt} > m_b/\sqrt{\xi}$, away from the $Z_2$-symmetric relation, while the twin top is always heavier. In summary, we conclude that $m_{\qt}/\Lqcdt \gtrsim 4$ and that $t_{\textrm{break}}$ is very large, much longer, as we show next, than the time it takes for the string to annihilate, which is the other possible fate of the string.%
\footnote{Heavier twin quarks, up to $m_{\qt} \approx 103.5 \GeV$, should have also been pair produced at LEP2 through an off-shell photon or $Z$. However, being the cross sections of $O(g'^2)$ and the quirks losing a significant fraction of their energy through electromagnetic radiation (see below), we expect LEP did not have the required sensitivity. Besides, even though the twin quarks modify at one loop SM processes like dilepton pair production (this being the same effect we encoded in the $Y$-parameter for twin masses beyond LEP reach, see Section~\ref{indirect}), we checked the effect is below experimental uncertainties \cite{Schael:2013ita}. We note that twin-QCD dynamics could also affect such processes, e.g.~via $Z$-glueball mixing, however we expect the effects to be small; also, the decay $Z \to \gt \gt \gt$ is below current constraints on non-standard $Z$ decays, $\BR(Z \to X_{\mathrm{BSM}}) \lesssim 10^{-4}$.}

The typical annihilation time depends a priori on the annihilation rates of the different states of energy and angular momentum the string can be in. 
In practice a good proxy is to consider only the low-lying low angular momentum ($\ell = 0$) bound states, whose lifetime is proportional to the classical crossing time of the quirks (or length of the string), 
\beq
\label{string}
T \sim \frac{m_{\qt}}{\Lqcdt^2} \approx 3 \times 10^{-24} \, \textrm{sec} \, \left( \frac{m_{\qt}}{100 \GeV} \right) \bigg( \frac{5 \GeV}{\Lqcdt}\bigg)^2 \, ,
\eeq
\ie our twin strings are microscopic, $L \sim 0.1 \, \textrm{fm}$.
This approach might naively seem at odds with the fact that most twin quarks will be produced relativistic, $\sqrt{\hat s} - 2 m_{\qt} \sim m_{\qt}$, therefore in highly excited states. However excited bound states have annihilation rates suppressed by their large angular momentum \cite{Kang:2008ea}. 
Therefore the low-lying states are eventually reached after the quirks radiate away most of their energy into (relatively) soft twin glueballs and photons \cite{Harnik:2008ax}. For the latter the typical radiation time can be estimated as $t_{\textrm{rad}}^{\textrm{QED}} \sim (3/8 \pi q_{\qt}^2 \alpha) m_{\qt}^3/\Lqcdt^4 \approx (2/q_{\qt}^2) \times 10^{-22} \, \textrm{sec}$ for the same parameters as in (\ref{string}) and $q_{\qt}$ the twin electric charge. This means that the de-excitation process is fast, in particular no displaced vertices in quirk pair-production and annihilation should be expected. Besides, twin glueball emission is also expected to contribute to the process, at least for energies where the kinematical suppression from the non-zero glueball mass is irrelevant.%
\footnote{Here we follow \cite{Kang:2008ea,Burdman:2008ek} where it is argued that the energetic strings quickly acquire large values of $\ell$, thus suppressing annihilation and favoring the process of de-excitation by glueball or photon emission. If this were not the case and annihilation took place before reaching the lowest bound states, the twin string would be best described as a broad resonance.}

Let us focus therefore on the lightest twin quarkonium bound states with $\ell = 0$, which are a pseudoscalar $\eta^{-+}$ and a vector $\Upsilon^{--}$, both electrically neutral (see also \cite{Cheng:2016uqk,Li:2017xyf}).%
\footnote{As usual the parity and charge-conjugation properties of the bound states are given by $P=(-1)^{\ell+1}$ and $C=(-1)^{\ell+s}$ respectively, $s$ being the spin.} 
Their mass is given to a good approximation by $m_{\eta,\Upsilon} \approx 2 m_{\qt}$, while their decay rates can be found e.g.~in \cite{Barger:1987xg,Cheung:2008ke,Fok:2011yc}.
Of special significance is the fact that the decay of $\Upsilon^{--}$ to a pair of twin gluons is forbidden, thus enhancing its branching ratio to SM final states. Neglecting the masses of the daughter particles as well as $m_Z$,
\begin{align}
\label{}
& \Gamma_{\Upsilon^{--} \to \gamma h, \, Z h} \simeq \frac{\alpha q_{\qt}^2 \yt_q^2 |\psi(0)|^2}{\pi m_{\Upsilon}^2} \left\{1,\, t_{\theta_W}^2 \right\} \, , \quad 
\Gamma_{\Upsilon^{--} \to f \bar f} \simeq \frac{4 N_c \alpha^2 [(Q_{V}^{\Upsilon f})^2 + (Q_{A}^{\Upsilon f})^2] |\psi(0)|^2}{m_{\Upsilon}^2} \, , \nn \\
& \Gamma_{\Upsilon^{--} \to \gt \gt \gt} \simeq \frac{40 \tilde \alpha_s^3 (\pi^2 - 9) |\psi(0)|^2}{81 \pi m_{\Upsilon}^2} \, ,
\label{Udecay}
\end{align}
where $Q_{V}^{\Upsilon f} = q_{\qt} (q_{f} + q_{Zf}^V/c_{\theta_W}^2)$ and $Q_{A}^{\Upsilon f} = q_{\qt} \, q_{Zf}^A/c_{\theta_W}^2$ with $q_i$ the electric charge of $i$ and $q_{Zf}^{V,A}$ the vector and axial $Z$-charges of the SM fermions, e.g.~for $f = e$, $q_{Ze}^{V} = \frac{1}{2} (-\frac{1}{2} + 2 s_{\theta_W}^2)$ and $q_{Ze}^{A} = - \frac{1}{4}$. The twin-QCD structure constant $\tilde \alpha_s$ is to be evaluated at $m_\Upsilon$,%
\footnote{To extract $\tilde \alpha_s$, we match it to the QCD coupling constant $g_s$ at $m_* = 5 \TeV$, \ie $\gt_s(5 \TeV) = g_s(5 \TeV)$, to then run it down to the relevant scale, taking into account that only one generation of twin quarks contributes to the running (and the twin top threshold).}
and $\psi(0)$ is the radial wave function of the bound state at the origin. For a $\Upsilon^{--}$ made of twin bottoms ($q_{\qt} = - \frac{1}{3}$), the decay to three twin gluons is the largest for all twin quarks masses of interest, $m_{\qt} \in (0.1, 1) \TeV$, while for a bound state of twin tops ($q_{\qt} = \frac{2}{3}$), decays to $\gamma h$ and $Z h$ dominate, owing to a large twin Yukawa coupling. Importantly, decays to SM fermions are always non-negligible, e.g.~$\BR(\Upsilon^{--} \to e^+ e^-) \approx 4 - 10 \%$, where the lower end is found for a twin top, quite independently of its mass.
In contrast, the $\eta^{-+}$ decays almost exclusively to two twin gluons, while its other allowed decay channels, to a pair of SM neutral gauge bosons (\ie $\gamma \gamma, ZZ, Z\gamma$), are suppressed by the small electromagnetic coupling constant, e.g.
\beq
\label{etadecay}
\Gamma_{\eta^{-+} \to \gt \gt} = \frac{8 \tilde \alpha_s^2 |\psi(0)|^2}{3 m_{\eta}^2} \, , \quad 
\Gamma_{\eta^{-+} \to \gamma \bar \gamma} = \frac{12  \alpha^2 q_{\qt}^4 |\psi(0)|^2}{m_{\eta}^2} \, .
\eeq
For the relevant range of quirk masses $\BR(\eta^{-+} \to \gamma \gamma) \approx q_{\qt}^4 (1 - 4) \%$, which means that even for a twin-top bound state the branching ratio is always below the per cent.

\begin{figure}[!t]
\begin{center}
\includegraphics[width=3.2in]{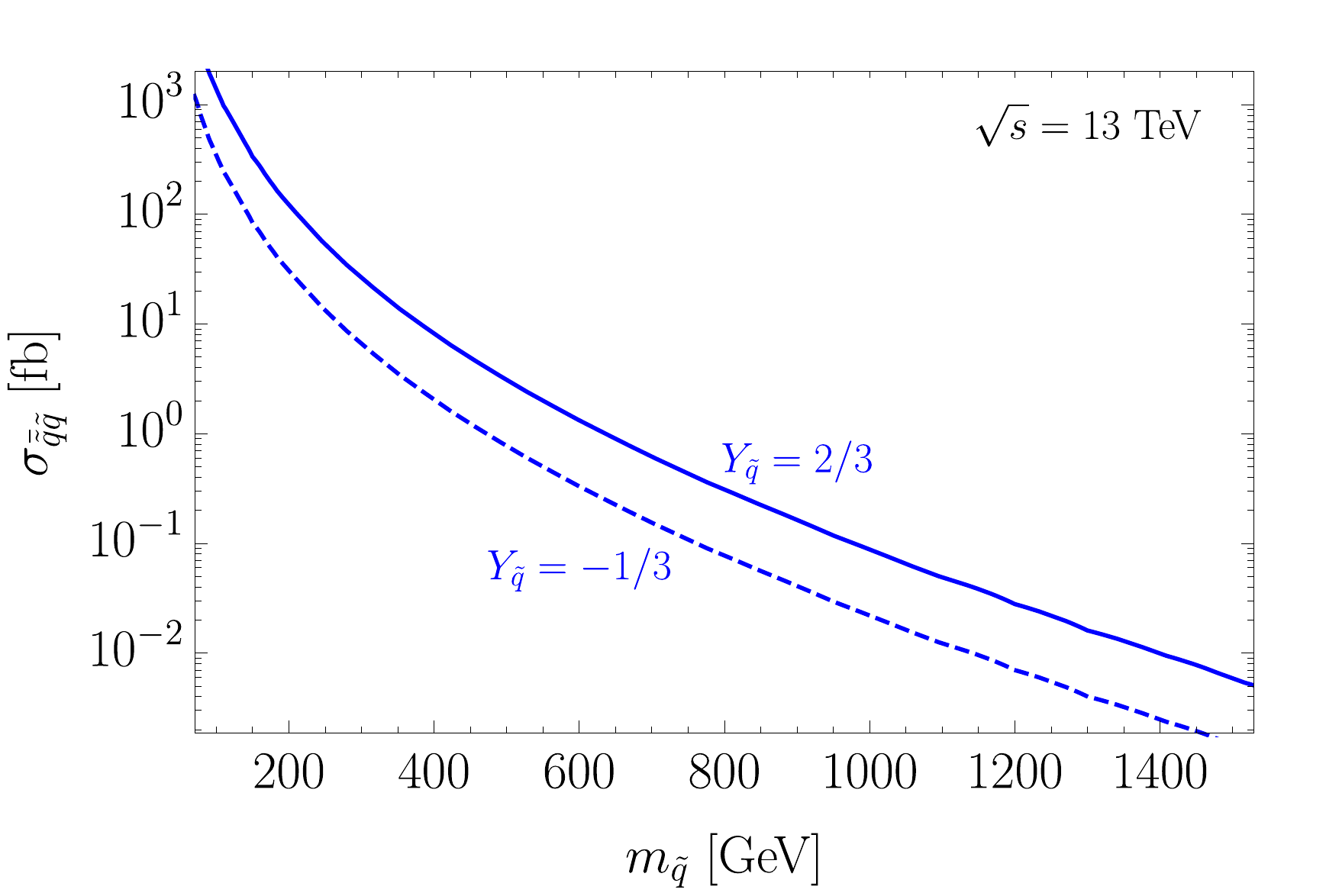}
\hspace{-1mm}
\includegraphics[width=3.2in]{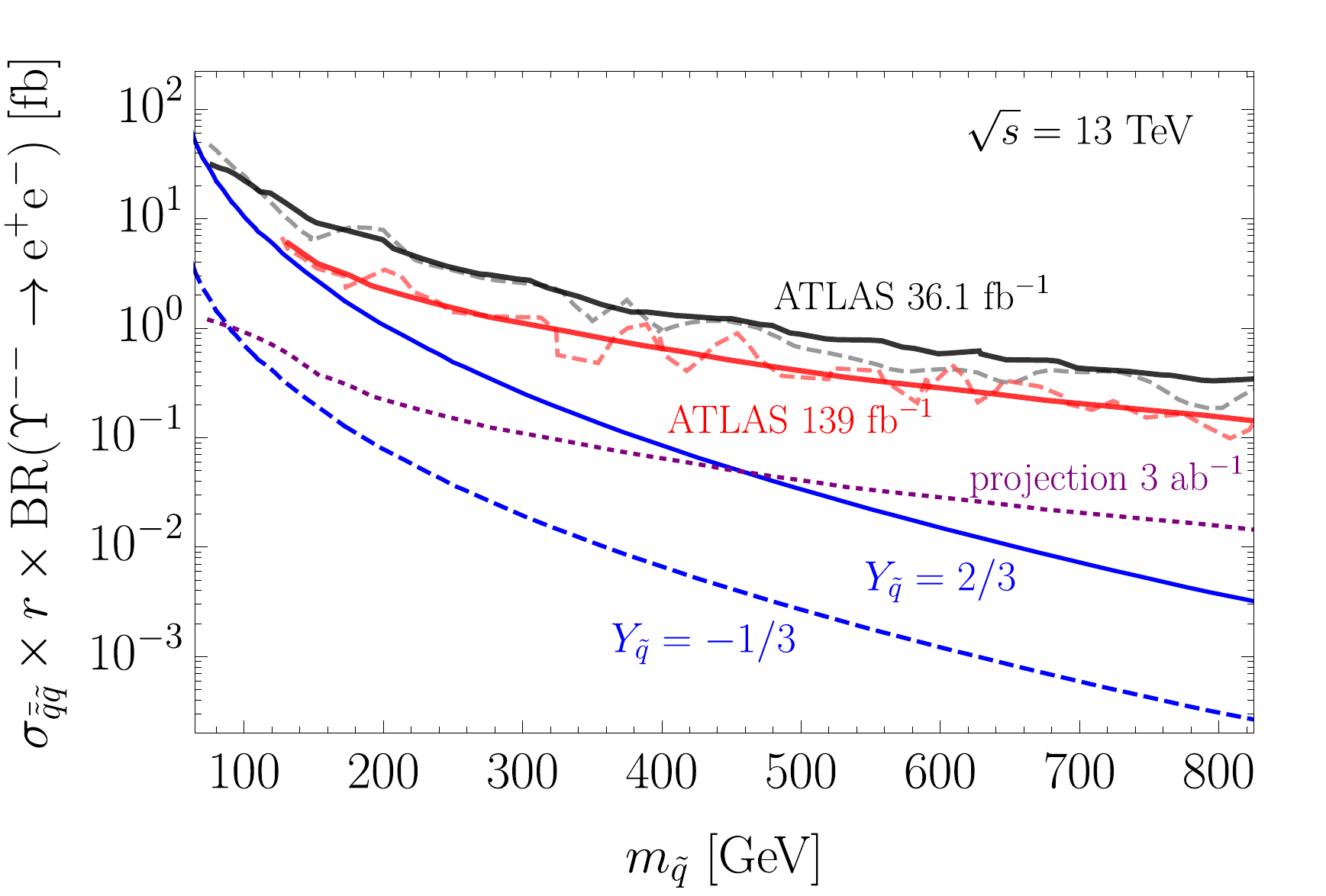}
\caption{(Left) Pair production cross sections of twin quarks with hypercharge $Y_{\qt} = \frac{2}{3}$ (solid) or $Y_{\qt} = -\frac{1}{3}$ (dashed). (Right) Upper $95\%$ CL limit on $Z'$ cross section times branching ratio to dileptons from ATLAS searches in $36.1 \fb^{-1}$ \cite{Aaboud:2017buh} (black solid) and in $139 \fb^{-1}$ \cite{Aad:2019fac} (red solid) of data at $\sqrt{s} = 13 \TeV$, where the $Z'$ is identified with the bound state $\Upsilon^{--} \sim \bar \qt \qt$ of mass $m_{\Upsilon} = 2 m_{\qt}$, with theory predictions for $Y_{\qt} = \frac{2}{3}$ (blue solid) and $Y_{\qt} = -\frac{1}{3}$ (blue dashed). Also shown the projected bound after $3 \ab^{-1}$ of integrated luminosity (see text for details).}
\label{fig:tquark}
\end{center}
\end{figure}

The substantial branching ratio of the $\Upsilon^{--}$ to dileptons make its production at the LHC one of the most promising avenues for detection of the twin quarks. To assess the sensitivity of current LHC searches for dilepton resonances, we plot in \Fig{fig:tquark} (left panel) the production cross section of pairs of twin quarks with hypercharge either as that of the twin top, $Y_{\qt} = \frac{2}{3}$, or as the twin bottom, $Y_{\qt} = -\frac{1}{3}$ (recall $q_{\qt} = Y_{\qt}$). We keep the mass $m_{\qt}$ a free parameter, but we recall that in our \tH model $m_{\topt} \simeq m_t /\sqrt{\xi}$ while $m_{\bt}$ only needs to be large enough for $Z \to \bar \bt \bt$ to be kinematically forbiden, but otherwise it is not bounded by naturalness considerations; we effectively vary $m_{\qt}$ by changing the corresponding vector-like mass $\widetilde m_{q}$ (\ie keeping the Yukawa coupling $\yt_q$ to its $Z_2$-symmetric value). 
The fraction of events that go through the formation and decay of $\Upsilon^{--}$ bound states rather than of $\eta^{+-}$ is expected to depend on their total decay rates, $\Gamma_{\Upsilon,\eta}$, as $r = 3 \Gamma_{\Upsilon}/(3 \Gamma_{\Upsilon}+\Gamma_{\eta})$ \cite{Cheng:2018gvu}, where the factor of three accounts for the number of degrees of freedom in the vector bound state versus in the pseudoscalar. Because of the large decay rate of the $\eta^{+-}$ to $\gt \gt$, we find $r$ is always below $25 \%$ for a twin-top bound state and below $5 \%$ for the twin bottom.
With these results we have computed the cross section for production and decay to $e^+ e^-$ (and $\mu^+ \mu^-$) of the $\Upsilon^{--}$, which we show in \Fig{fig:tquark} (right panel) for a bound state made of twin quarks with either top-like of bottom-like hypercharges. 
Comparing them with the latest ATLAS bounds on a $Z'$ decaying to dileptons \cite{Aad:2019fac}, we find that the current $139 \fb^{-1}$ of data at $\sqrt{s}= 13 \TeV$ are not quite enough to probe the small cross sections associated with the hypercharged quirks. However, the order of magnitude increase in sensitivity needed to probe the relevant twin-top cross sections could potentially be achieved at the LHC with $3 \ab^{-1}$ of integrated luminosity, as we also shown in \Fig{fig:tquark}. Such a expected a limit has been obtained from a naive luminosity rescaling, after contrasting the improvement of the expected bounds from $36 \fb^{-1}$ \cite{Aaboud:2017buh} to $139 \fb^{-1}$ \cite{Aad:2019fac} of data (note that the HL-LHC projections in \cite{CidVidal:2018eel} do not go below $2 \TeV$ resonance masses).
In addition, searches for $\gamma h$ resonances are also relevant in view of the large branching ratios of the twin-top $\Upsilon^{--}$ to this final state. Current analyses \cite{Aaboud:2018fgi} set a constraint $\sigma \cdot \BR \lesssim 10 \fb$, roughly one order of magnitude above the cross section predicted in our scenario for $m_{\Upsilon} \sim 1 \TeV$, which makes this a very interesting signature for the high-luminosity phase of the LHC.

Similarly, the cross section for production and decay to diphotons of the pseudoscalar, $\sigma_{\bar \qt \qt} \cdot (1-r) \cdot \BR(\eta^{-+} \to \gamma \gamma) \approx 1.6 \times 10^{-2} \fb$ for $m_{\eta} = 1 \TeV$ and $Y_{\qt} = \frac{2}{3}$, is for the most part down by one/two orders of magnitude compared to current LHC bounds on diphoton resonances \cite{Aaboud:2017yyg}. Once again the interesting conclusion is that the HL-LHC should have enough sensitivity to probe a twin-top bound state decaying to $\gamma\gamma$ of mass as expected from twin parity, therefore reaching the relevant parameter space.
We note that the situation is certainly different if twin quarks had larger hypercharges, for instance if $Y_{\qt} = 2$ current diphoton searches would already exclude $m_{\qt} \lesssim 1 \TeV$.

\subsubsection{Twin glueballs} \label{glueballs}

In our exceptional twin Higgs, twin glueballs are the lightest states of the twin sector. They are produced from Higgs decays, as in other scenarios of neutral naturalness \cite{Craig:2015pha,Cohen:2018mgv}, as well as from the decays of the twin-quark bound states, as discussed in the previous section.%
\footnote{Another potential source of glueballs is from the previous deexcitation of the quirks, which we neglect here since it is unclear if such radiation would dominate over the electromagnetic one.} 
While the former process dominates for heavy twin quarks, it is interesting that the decays of twin-bottom bound states lighter than approximately $300 \GeV$ dominate the inclusive glueball production at $\sqrt{s} = 13 \TeV$: $\sigma^{(h)}_{\gt} \approx 67 \fb \, (\xi/0.1)^2$ while $\sigma^{(\Upsilon+\eta)}_{\gt}$ is very near the total quirk production cross section, shown on the left panel of \Fig{fig:tquark}, since bound-state annihilation to glueballs always dominates. 

The mass, lifetime and decay modes of a given glueball are determined by its quantum numbers, see e.g.~\cite{Juknevich:2009ji,Juknevich:2009gg}. Therefore, the actual signatures associated to glueball production depend on which and how many glueballs are produced from the aforementioned decays. These questions however cannot be reliably answered without a proper understanding of the non-perturbative twin-color dynamics (see however~\cite{Lichtenstein:2018kno}). We will therefore focus our attention on the lightest glueball $0^{++}$, whose dominant decays are mediated by the Higgs, via the dimension-6 operator $|H|^2 \widetilde G_{\mu \nu} \widetilde G^{\mu \nu}$, as well as on the lightest glueball for which such decays are forbidden, i.e.~the $0^{-+}$, which decays almost exclusively to two photons. Both of these glueballs, of mass $m_{0^{++}} \equiv m_0 \approx 6.9 \Lqcdt$ and $m_{0^{-+}} \approx 1.5 m_0$, are expected to represent a significant low-energy population of the energetic twin gluons originally produced, with the $0^{++}$ expected to dominate if the kinetic energy of the twin gluons is low, being the lightest. Other glueballs could also yield interesting and somewhat distinct phenomenology, however the $0^{++}$ and $0^{-+}$ are good representatives of the particularities of our exceptional twin Higgs with respect to previous models. For instance, we checked that the $2^{++}$ glueball decays predominantly to diphotons and is stable on collider scales, as the $0^{-+}$, see \Eq{ffgg}.

\begin{figure}[!t]
\begin{center}
\includegraphics[width=2.7in]{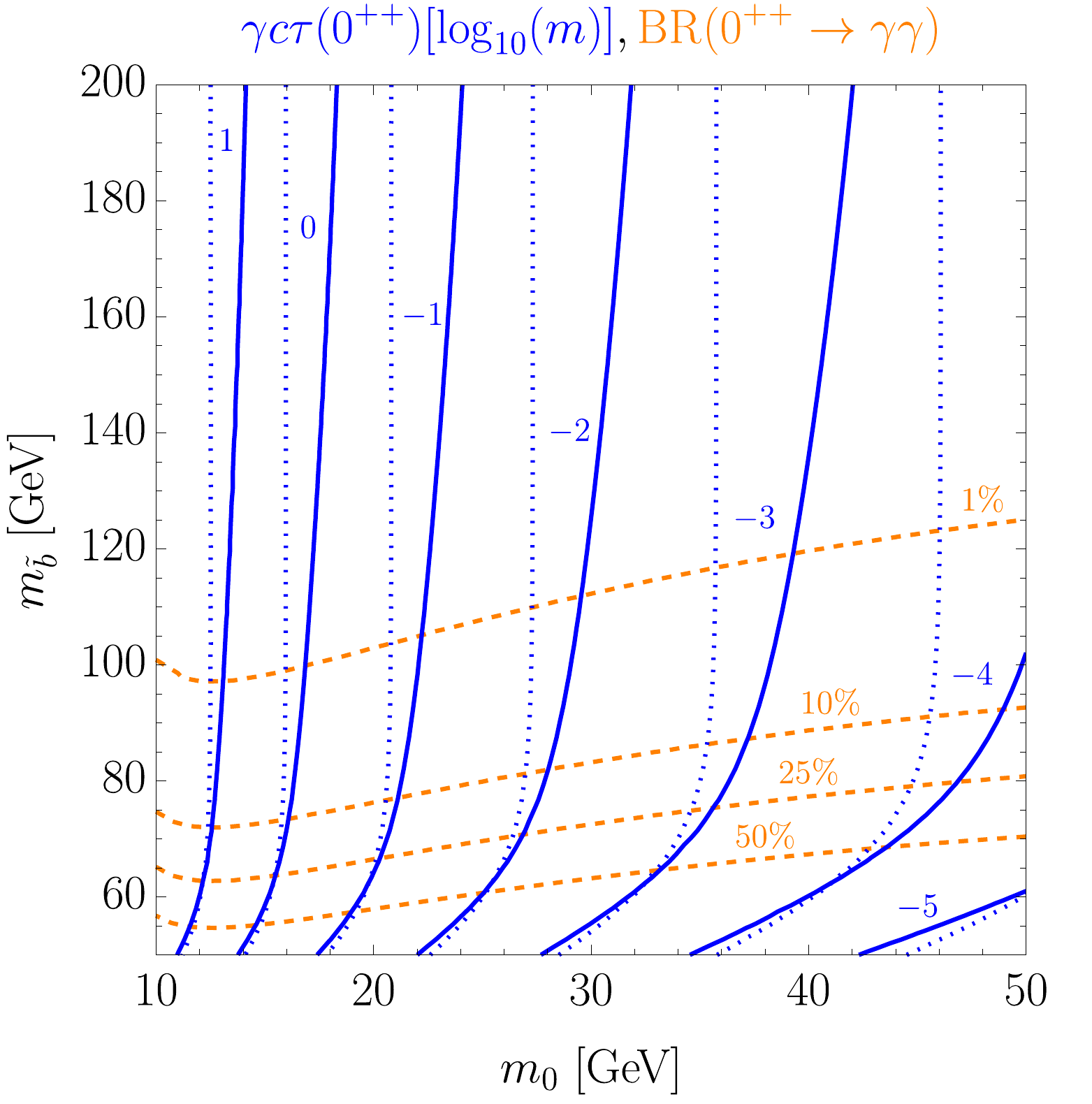}
\hspace{7mm}
\includegraphics[width=2.7in]{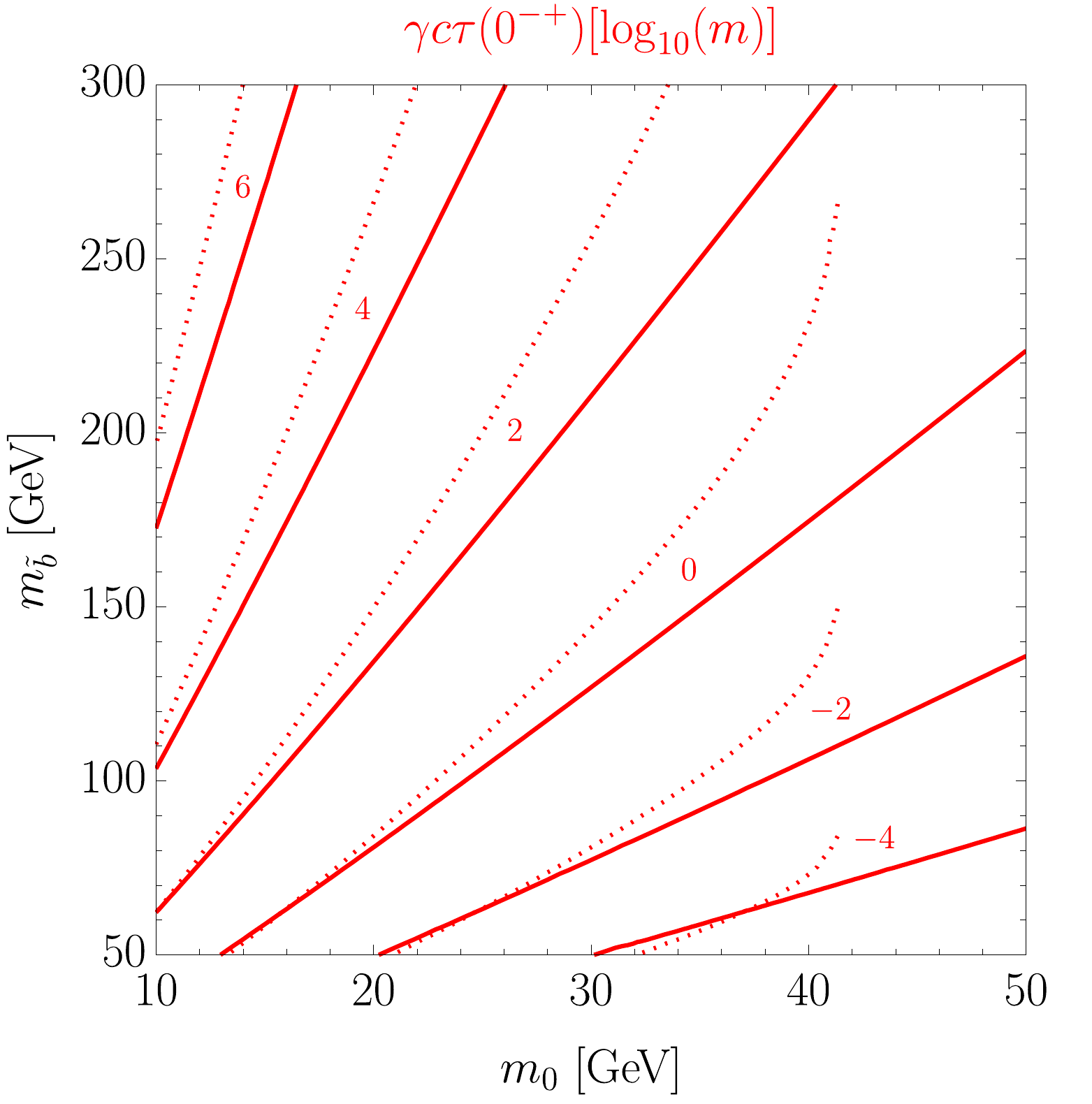}
\caption{(Left) Branching ratio of the lightest glueball $0^{++}$ to diphotons (orange) and lab decay length (blue) from production via either Higgs decays (dotted) or $\eta^{-+}$ decays (solid), for $\xi = 0.1$. (Right) Lab decay length of the lightest parity-odd glueball $0^{-+}$ (red), of mass $m_{0^{-+}} \approx 1.5 \, m_{0}$, from production via either Higgs decays (dotted) or $\eta^{-+}$ decays (solid). The decay lengths have been computed under the assumption that the parent particle produces two glueballs only, thus they should be understood as upper bounds.}
\label{fig:glueball}
\end{center}
\end{figure}

Significant branching ratios to diphotons is in fact the identifying feature of the glueballs in our scenario. These arise via dimension-8 operators of the form $c \, F_{\mu \nu}F^{\mu \nu} \widetilde G_{\rho \sigma} \widetilde G^{\rho \sigma}$ generated by a twin-quark loop, of size
\beq
\label{cffgg}
c \sim \frac{\alpha \, \tilde \alpha_s Y_{\qt}^2}{m_{\qt}^4}\,,
\eeq
which is therefore enhanced at low twin-quark masses. For this reason in the following we will consider the twin bottom as a key player in glueball phenomenology, and comment when relevant on the differences that would arise if the $\bt$ is decoupled and only the twin top contributes, e.g.~\Eq{cffgg} would be enhanced due to the larger hypercharge of $\topt$ but suppressed by its larger mass. In the left panel of \Fig{fig:glueball} we show the branching fraction to diphotons of the lightest glueball in the $(m_0,m_{\bt})$ plane, in the range of glueball mass expected from our estimates of $\Lqcdt$. As anticipated, only for light twin bottoms this channel dominates over the standard Higgs-mediated decay to a pair of SM bottoms. The decay length of the $0^{++}$ can be approximated, for sufficiently large $m_{\bt}$ and/or $\xi$ by $c \tau_{0^{++}} \approx (0.4 \, \mathrm{m}) (15 \GeV/m_0)^7 (0.1/\xi)^2$ as in standard \tH constructions. Therefore, such glueballs are relativity long-lived, giving rise to displaced vertices in a significant fraction of the parameter space. This decay length, but in the lab frame, is also shown in the left panel of \Fig{fig:glueball}, where the boost factor is either from the decay of a Higgs (blue dotted) or the decay of the twin-bottom bound states (blue solid), to two, and only two, glueballs. More realistically, such decays will give rise to more glueballs, meaning these lengths should be considered as upper bounds.
One of the new and exciting features of our glueballs is found in the right panel of \Fig{fig:glueball}, where we show the lab decay length of the $0^{-+}$, which almost exclusively decays to $\gamma \gamma$ and it is very long-lived, in fact stable on collider scales in a large portion of parameter space. This can be easily understood from the fact that its rest-frame decay length scales as
\beq
\label{ffgg}
c \tau_{0^{-+}} \approx (55 \, \mathrm{m}) \left(\frac{m_{\bt}}{150 \GeV}\right)^8 \left(\frac{20 \GeV}{m_0}\right)^9 \left(\frac{-1/3}{Y_{\bt}}\right)^4 \,,
\eeq
which in the lab frame translates to up to $250 \, \mathrm{m}$.
Since the dependence on either $m_0$ or $m_{\bt}$ is very strong, these glueballs could give rise either to displaced photons or to missing energy signatures in the detector. In the latter case, the glueballs would eventually decay to a pair of photons, which could potentially be detected by one of the proposed detectors dedicated to long-lived particle searches \cite{Curtin:2018mvb,Gligorov:2017nwh,Gligorov:2018vkc}.
Besides, notice that if the twin bottom is decoupled, $c \tau_{0^{-+}}$ would be substantially larger, since the decay would only be mediated by the heavier twin top.

\begin{figure}[!t]
\begin{center}
\includegraphics[width=2.7in]{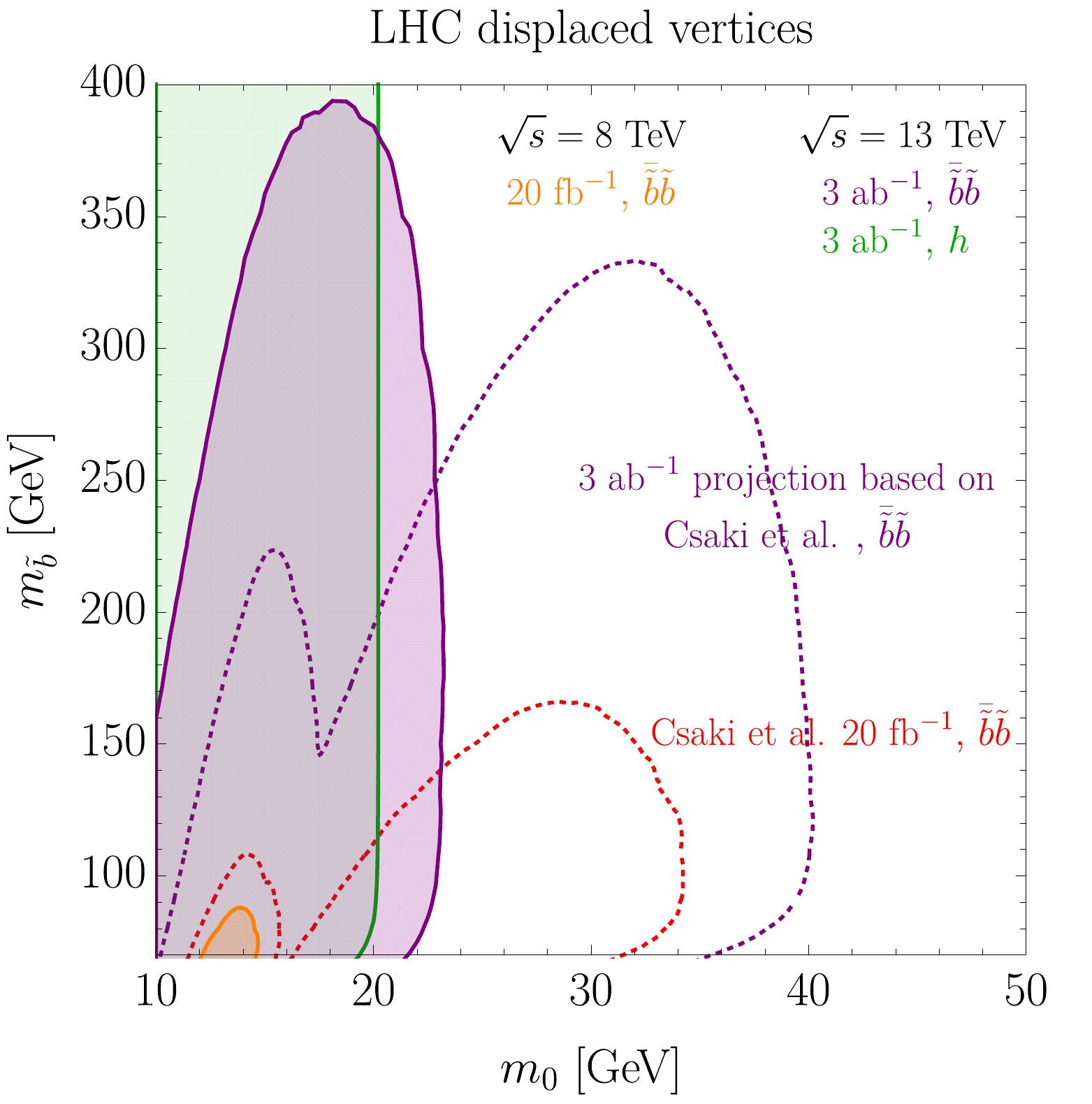}
\hspace{7mm}
\includegraphics[width=2.7in]{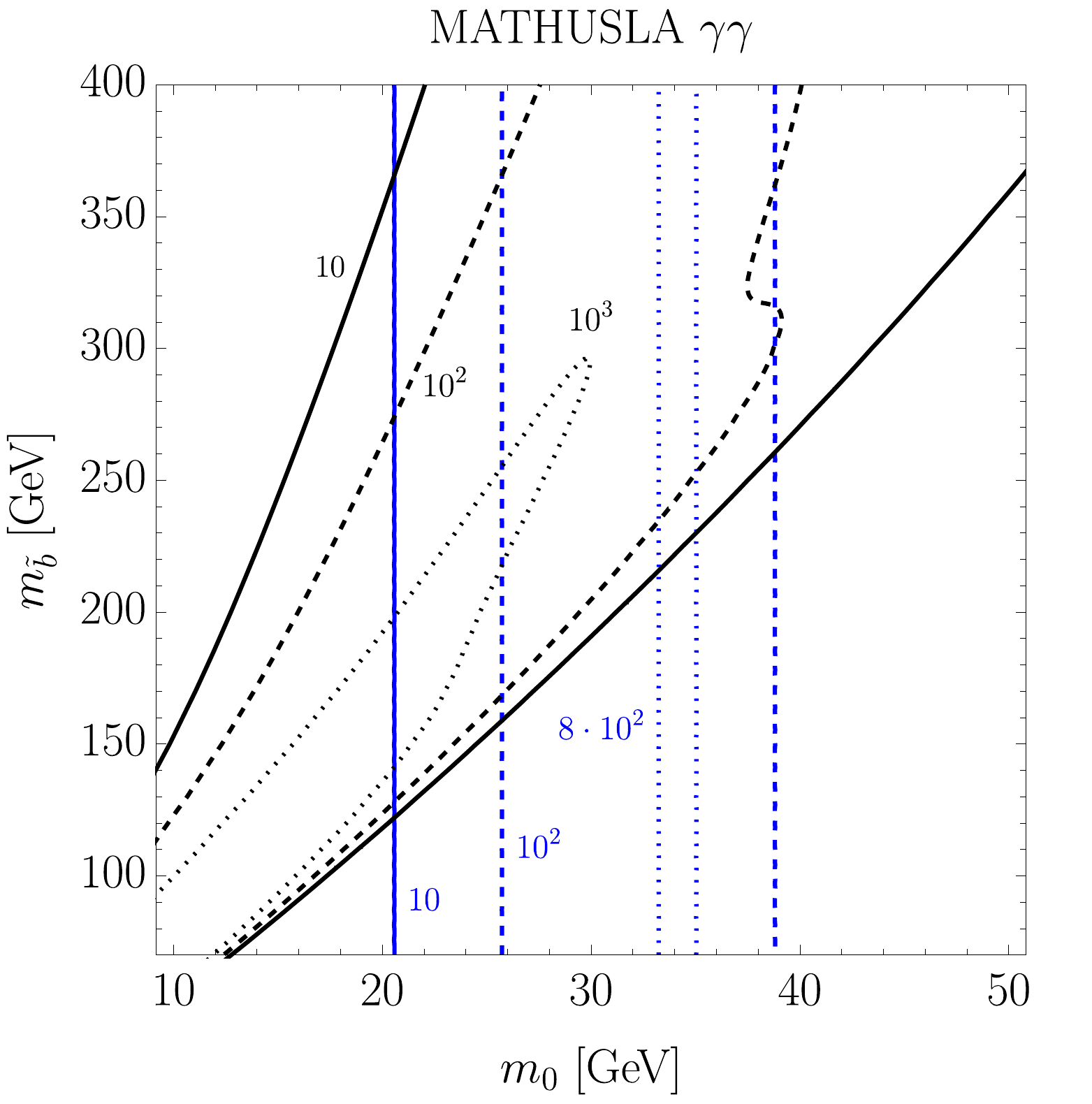}
\caption{(Left) Current and projected LHC constraints on the $0^{++}$ glueball mass $m_0$ and the twin-bottom mass $m_{\bt}$ from searches for displaced vertices, for $\xi = 0.1$. The bounds shown correspond to $0^{++}$ production from annihilation of twin-bottom bound states at $8 \TeV$ in $20 \fb^{-1}$ of ATLAS data \cite{Aad:2015uaa} (orange), and projected searches from \cite{Csaki:2015fba} at $13 \TeV$ with $20 \fb^{-1}$ (dotted red) and $3 \ab^{-1}$ (dotted purple), as well as from the HL-LHC \cite{Cepeda:2019klc} (purple). Shown also the HL-LHC bounds on $0^{++}$ production from Higgs decays (green). (Right) Expected number of glueballs decaying to diphotons inside the MATHUSLA detector, from twin gluons produced at the $13 \TeV$ LHC after $3 \ab^{-1}$ of integrated luminosity, via decays of the Higgs and twin-bottom bound-states (black) or Higgs and twin-top bound-states with $m_{\topt} \simeq m_t/\sqrt{\xi}$ (blue), for $\xi = 0.1$.}
\label{fig:gluebound}
\end{center}
\end{figure}

Before discussing the experimental prospects for such collider-stable glueballs, let us understand the sensitivity of current and future LHC data to the $0^{++}$ glueballs. Since these are long-lived but always decay within the detector, they give rise to displaced vertices. This kind of signals have been studied in detail in several works, e.g.~\cite{Curtin:2015fna,Csaki:2015fba,Chacko:2015fbc,Coccaro:2016lnz,Kilic:2018sew}, and we reinterpret here some of their results to account for the effects of our hypercharged twin quarks.
In the left panel of \Fig{fig:gluebound} we show the constraints on $m_0$ and $m_{\bt}$ from several searches for displaced vertices.
These in general focus on decays either in the inner detector, $c \tau \gamma \lesssim 25 \, \mathrm{cm}$ ($\gamma$ the boost factor), therefore sensitive to heavier $0^{++}$ glueballs, or in the hadronic calorimeter and/or muon spectrometer, that is $2 \, \mathrm{m} \lesssim c \tau \gamma \lesssim 10 \, \mathrm{m}$ and thus of relevance for lighter glueballs. 
The current most sensitive experimental analysis that is well suited to the features of our twin glueballs is from ATLAS at $8 \TeV$ and with $20 \fb^{-1}$ of data \cite{Aad:2015uaa} (small orange region in \Fig{fig:gluebound}). While this search barely reaches the interesting region, the theory projections of \cite{Csaki:2015fba} for $13 \TeV$ and $20 \fb^{-1}$ (dotted red line) show that with current data one could already probe twin-bottom masses up to $150 \GeV$, at least for glueballs that decay within the inner detector.%
\footnote{Even though most strategies, and in particular those of \cite{Csaki:2015fba}, focus on displaced vertices from Higgs decays, we naively assume that similar sensitivities can be achieved for glueballs from twin-quark bound-state decays.}
We should note however that these projections are somewhat uncertain, in particular to date there is no experimental analysis demonstrating the sensitivity to such heavy glueballs, while the region at small $m_0$, which overlaps with the ATLAS $8 \TeV$ analysis \cite{Aad:2015uaa}, has not been corroborated by a recent $13 \TeV$ analysis that studies glueballs decaying in the hadronic or the outer edge of the electromagnetic calorimeters \cite{Aaboud:2019opc}. We have checked that this experimental analysis, based on $10.8 \fb^{-1}$ of data, does not yield any constraint on out parameter space. Besides, the analysis of \cite{Csaki:2015fba} is not sensitive to glueballs originating from Higgs decays (for $\xi \lesssim 0.1$), but only to those from the annihilation of twin-bottom bound states. In \Fig{fig:gluebound} we also show the region covered by a naive (by $1/\sqrt{L}$) rescaling of the projections of \cite{Csaki:2015fba} to $L = 3 \ab^{-1}$ (dotted purple). This simple extrapolation shows that the high-luminosity phase of the LHC could considerably extend the coverage up to $m_{\bt} \sim 300 \GeV$. 
Moreover, we have also recasted the HL-LHC projections from \cite{Cepeda:2019klc}, which shows sensitivity to light $0^{++}$ glueballs produced from twin-bottomonium annihilation up to $m_{\bt} \sim 400 \GeV$ (purple region), as well as glueballs from Higgs decays (green). The latter region is independent of $m_{\bt}$ since the twin bottom does not contribute significantly to the Higgs branching ratio into twin gluons, being instead solely determined by $\xi$. Indeed, for $\xi \lesssim 0.01$ the green region in \Fig{fig:gluebound} disappears. In addition, both the constraints associated to Higgs and twin quarkonium annihilation extend to higher $m_0$ values for smaller $\xi$, since the decay rate of the $0^{++}$, dominated by Higgs exchange, decreases. Let us also comment on the dependence of these constraints on the electric charge of the twin quarks, in particular for $Y_{\topt} = 2/3$ as it corresponds to the twin top. A larger $Y_{\qt}$ implies a larger production cross section of the twin quarks for the same mass, thus the main effect is that the constraints from searches for displaced vertices extend to heavier $m_{\qt}$. We find that $3 \ab^{-1}$ could probe up to $m_{\topt} \sim 520 \GeV$ for light glueballs or $m_{\topt} \sim 450 \GeV$ for heavy ones, close to the twin-top mass expected from twin parity; we could forecast that a dedicated search would achieve the required sensitivity.
Let us add that in the above we have always assumed that two, and only two, $0^{++}$ glueballs are produced during twin-color glueballization.

The very slow decay of twin glueballs such as the $0^{-+}$, see \Eq{ffgg}, could potentially be detected by a surface detector such as MATHUSLA \cite{Curtin:2018mvb}, granted its efficiency to identify photons is not negligible. In the right panel of \Fig{fig:gluebound} we show a (very) rough estimate of the number of photons that would decay inside the detector: $L \cdot \sigma_{0^{-+}} \cdot P(d)$, with $L = 3 \ab^{-1}$, the $13 \TeV$ cross section $\sigma_{0^{-+}}$ includes glueballs from twin gluons produced in both Higgs ($\xi = 0.1$) and twin-bottomonium decays, and $P(d) = e^{-d_i/d} (1-e^{-(d_i+\Delta d)/d}) \Omega_M/4 \pi$ is our naive estimate of the probability for the glueball to decay inside MATHUSLA, where $d_i \approx 225 \, \mathrm{m}$ (distance from the interaction point to center of the closer horizontal edge of the detector), $\Delta d \approx 45 \, \mathrm{m}$ (the corresponding maximum distance to the farther edge of the detector), $d$ is to be identified with the lab-frame decay length of the glueball, $(c\tau\gamma)_{0^{-+}}$, and $\Omega_M \approx 0.3$ the solid angle covered by the detector.%
\footnote{A proper estimate would take into account that the distances to the closer and farther edges of the detector depend on the direction of incidence of the glueball.}
Once again we have assumed for simplicity that at least two, but only two, $0^{-+}$ glueballs arise from the hard-scattered twin gluons.
We therefore conclude that large numbers of glueballs, up to $O(10^3)$ after $3 \ab^{-1}$ of integrated luminosity, could potentially decay to diphotons within MATHUSLA. Interestingly, this conclusion does not depend on the presence of light twin bottoms in the spectrum, as shown by the blue contours in \Fig{fig:gluebound}, corresponding to glueballs from the decays of Higgs and twin-top bound states at their twin-symmetric mass ($\xi = 0.1$).

We conclude this section by noticing that another potentially relevant signature of our glueballs is given by the process $pp \to \gt \gt \to 0^{++} 0^{-+} + X$, where the $0^{++}$ decays displaced to $\gamma \gamma$ (even if this is a subleading decay channel, see \Fig{fig:glueball}), while $0^{-+}$ escapes the detector and thus shows up as missing energy \cite{Chatrchyan:2012jwg,Aad:2014gfa}.

\subsection{Fraternal $Z'$} \label{Zprime}

\begin{figure}[!t]
\begin{center}
\includegraphics[width=3.23in]{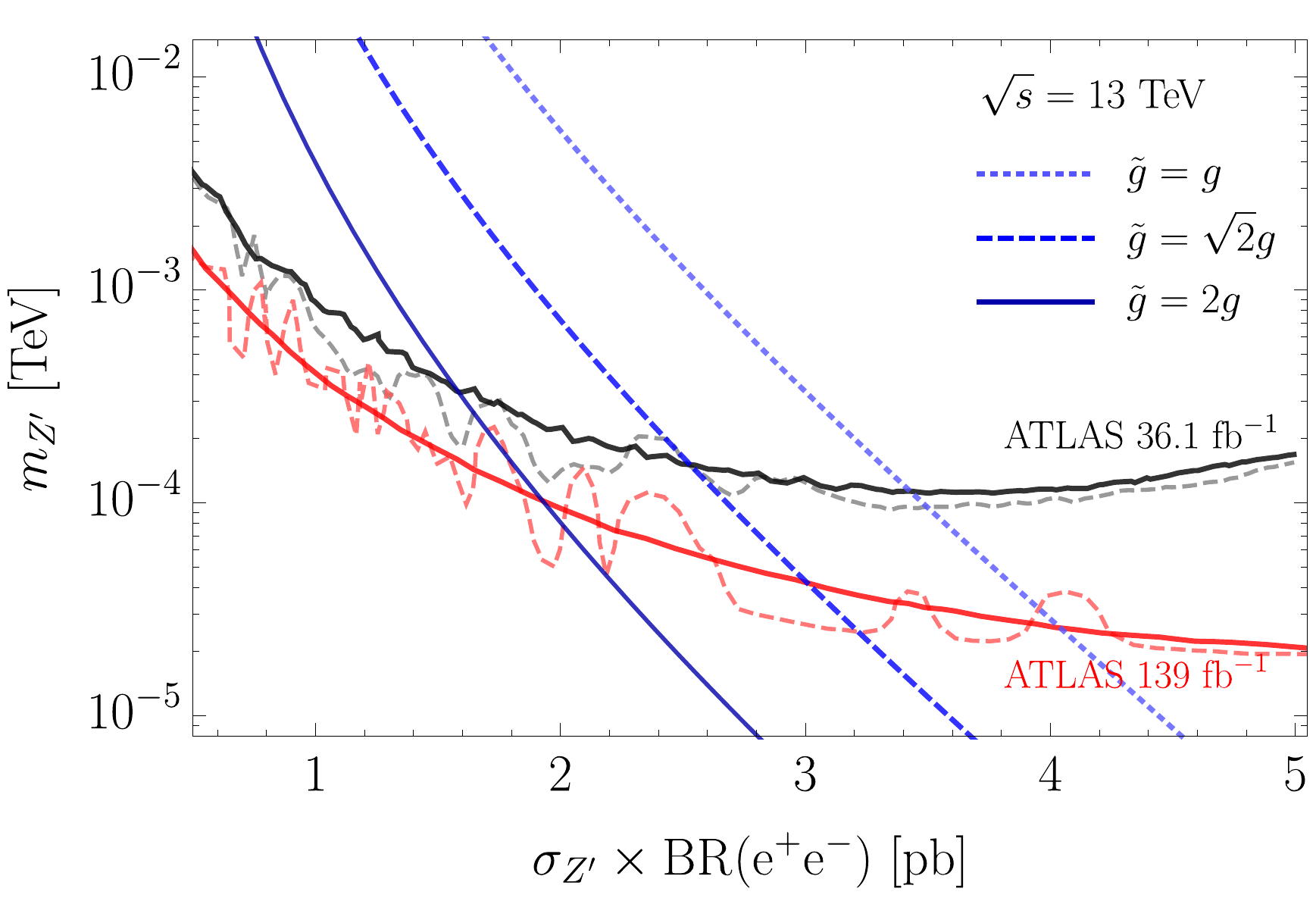}
\hspace{-1mm}
\includegraphics[width=3.18in]{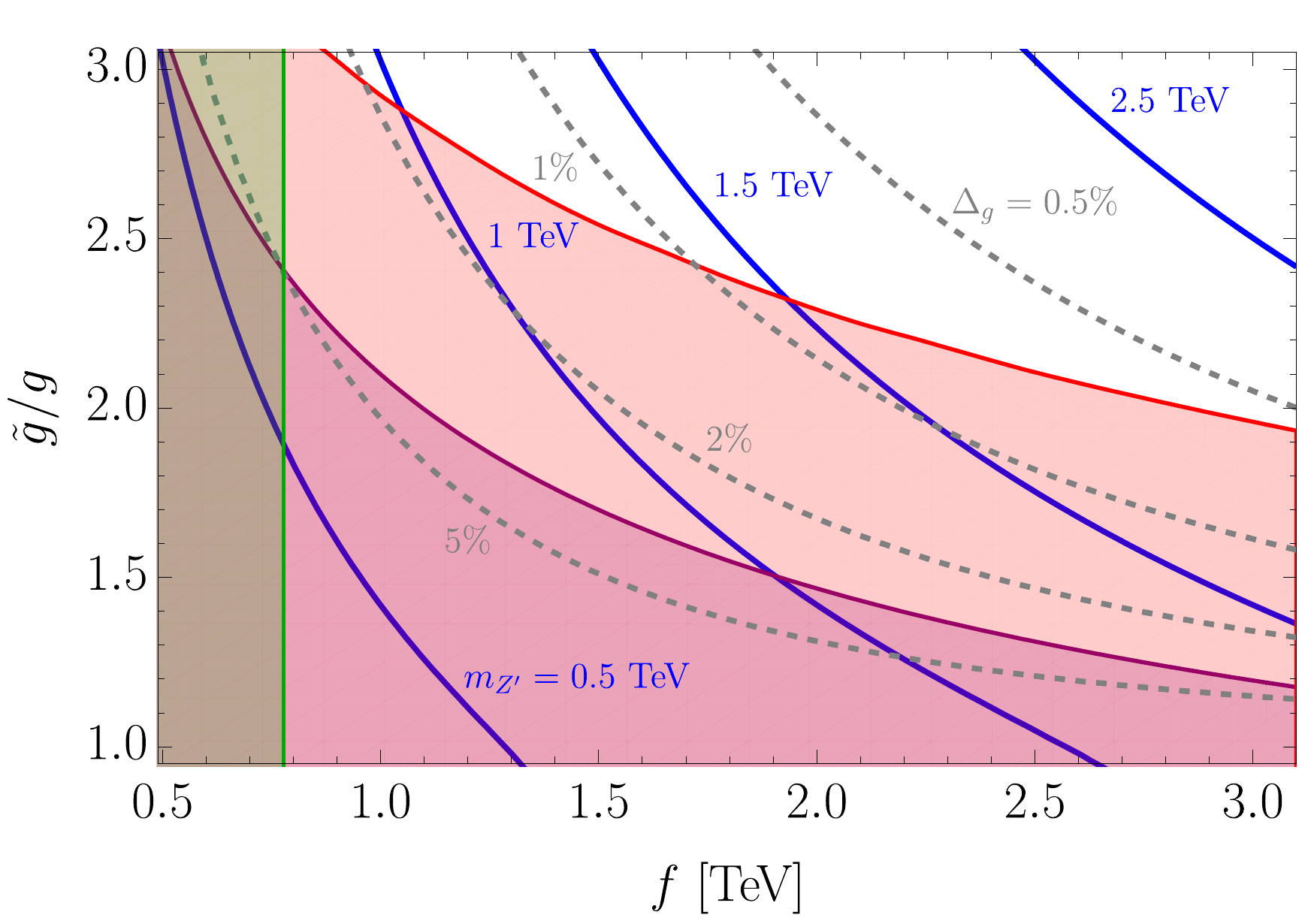}
\caption{(Left) Upper $95\%$ CL limits on $Z'$ cross section times branching ratio to dileptons as a function of the $Z'$ mass from a published search in $36.1 \fb^{-1}$ of data at $\sqrt{s}= 13 \TeV$ \cite{Aaboud:2017buh} (expected bound as black solid and observed as dashed) and from a more recent search with $139 \fb^{-1}$ \cite{Aad:2019fac} (red), as well as the prediction for the twin $Z'$ for different $SU(2)_{\Lt}$ gauge couplings: $\gt = g, \sqrt{2} g, 2 g$ (blue dotted, dashed, solid respectively). (Right) In the plane of Higgs decay constant $f$ and $\gt/g$, excluded regions from $Z'$ bounds (red shaded), the $Y$-parameter (purple shaded) and $\xi < 0.1$ (green shaded). Also shown contours of constant twin $Z'$ mass (blue solid) and Higgs tuning (grey dashed), the latter computed with $g_\rho = 6$.}
\label{fig:Zp}
\end{center}
\end{figure}

In the the fraternal model, the spontaneous breaking of $SU(2)_{\Lt} \times U(1)_{\Y}$ to $U(1)_Y$ gives rise to a tree-level contribution to $Y$-parameter, which decouples with both the twin gauge coupling $\gt$ and the symmetry-breaking scale $f$ as
\beq
Y_{Z'} = \frac{g^2 g'^2 \xi}{\gt^4 (1-\xi)} \simeq \frac{g'^2 }{\gt^2} \frac{m_W^2}{m_{Z'}^2} \, ,
\label{YZp}
\eeq
where on the r.h.s.~we neglected subleading terms in $\xi$ and $g'/\gt$. The constraint from LEP $Y \lesssim 1 \cdot 10^{-3}$ then leads to an important upper bound on the parameter combination $\gt^4 f^2$. Nevertheless, except for very large $\gt$ or unless some ad hoc assumption is made on the decays of the $Z'$, this constraint is milder than the one arising from direct $Z'$ searches at the LHC, as shown in the right panel of \Fig{fig:Zp}.

We recall that the twin $Z'$ is the axial combination of the $\Ah$ and $\Wt^3$ gauge bosons, with mass $m_{Z'} \sim \gt f /2$. Its couplings to the SM and twin fermions can be written as
\beq
\gt Z'^\mu \left( c_{\thetat} \mathcal{J}_\mu^{Q}(\psit) - \frac{1}{c_{\thetat}} \mathcal{J}_\mu^{\Y}(\psit) - \frac{s_{\thetat}^2}{c_{\thetat}} \mathcal{J}_\mu^{Y}(\psi) \right) + O(\xi) \,, 
\label{gZp}
\eeq
where $s_{\thetat} = g'/\gt$ and $\mathcal{J}^{Q,\Y,Y}$ are the $U(1)_{Q,\Y,Y}$ currents made of either SM fermions $\psi$ or twin fermions $\psit$, with twin parity enforcing $\Y(\psit_i) = Y(\psi_i)$ and $Q(\psit_i) = Q(\psi_i)$.
The $O(\xi)$ term stands for corrections induced after EWSB, which we safely neglected in the following.
From the last term in brackets one can immediately see that the $Z'$ can be produced from a SM fermion pair, e.g.~$q \bar q$ at the LHC, with a cross section that scales as $\sigma_{Z'} \sim g'^2 (g'/\gt)^2$ for fixed $Z'$ mass and $\gt \gg g'$. In this limit the width of the $Z'$ is dominated by decays to twin fermions, with branching ratios to SM fermions scaling as $\BR(\bar \psi \psi) \sim (g'/\gt)^2$. Therefore, for large values of $\gt$ both production and SM decays are suppressed. 

We find the strongest direct constraints on the $Z'$ arise from ATLAS searches for a resonant peak in dilepton ($e^+ e^-, \mu^+ \mu^-$) invariant mass distributions \cite{Aaboud:2017buh,Aad:2019fac}. The bounds on $\sigma \cdot \BR$ are shown in the left panel of \Fig{fig:Zp}, along with the theoretical prediction for three different values of the twin $SU(2)_{\Lt}$ gauge coupling and assuming one full generation of twin fermions the $Z'$ can decay to.%
\footnote{The constraints would be mildly weaker if three twin generations instead of one were to be considered.}
In the twin-symmetric case $\gt = g$, we find $m_{Z'} \gtrsim 4.1 \TeV$, a stringent bound that, according to \Eq{Zpmass}, implies $f \gtrsim 10 \TeV$, way beyond natural values. This motivates larger values of $\gt$, for which the constraints on $m_{Z'}$ are milder and the corresponding amount of fine-tuning is smaller. This is shown in the right panel of \Fig{fig:Zp}, where we plot the excluded regions from LHC direct searches and from the $Y$-parameter \Eq{YZp} in the ($f$, $\gt/g$) plane, along with contours of fixed $m_{Z'}$ and tuning $\Delta_g$ associated to the gauge contributions to the Higgs potential for $g_\rho = 6$; values of $\Delta_g \gtrsim 0.5 \%$ can be achieved for $\gt \gtrsim 2 g$.

In view of these results, the question arises of how $\gt$ could happen to be larger than $g$ at low energies. The naive possibility of decoupling two full twin generations at a high-energy scale above $m_*$ where $\gt \approx g$, is at odds with the requirement of approximately equal color and twin-color gauge couplings at $m_*$. Therefore, other ideas seem to be required (e.g.~large threshold corrections or extra $SU(2)_L$-charged matter). These difficulties nevertheless suggest that the easiest realization of our \tH scenario is the minimal one, in which the twin $SU(2)_{\Lt}$ gauge symmetry has been broken above $m_*$.


\section{Conclusions} \label{conclusion}

Twin Higgs models enjoy phenomenological signatures that, while linked to the naturalness of the electroweak scale, significantly differ from those of the standard TeV solutions of the hierarchy problem. Instead of collider signals associated with production of colored particles, e.g.~top partners, stops and gluinos, that subsequently decay to top/bottom and Higgs/EW gauge bosons or missing energy, in the \tH scenario the phenomenology at low energies is dominated by states that predominantly couple to the SM via the Higgs, thus giving rise for instance to exotic Higgs decays or heavy Higgs-like signatures. This kind of signals are more difficult to detect at colliders such as the LHC, which is the reason why current constrains are mild. More important is however the fact that large luminosities, expected to be delivered by the LHC in the forthcoming years, could potentially set the scales even.

In this paper, we have presented a new \tH construction that nicely exemplifies the exciting prospects for discovery of these unorthodox low-energy manifestations of theories solving the hierarchy problem, where the states that account for the little hierarchy between the Higgs mass and the cutoff $m_*$ of the Higgs sector are colorless but hypercharged. This is a consequence of the symmetry structure of the strong sector, whose dynamics is ultimately responsible for shielding the EW scale from UV thresholds while giving rise to a seemingly elementary Higgs in the IR. Besides a $Z_2$ parity that exchanges color and its twin, a global $SO(7)$ symmetry, spontaneously broken to the $G_2$, delivers a Higgs and a twin Higgs that carries hypercharge. These couple to the SM fermions and their twins, in particular to a twin top with $2/3$ of electric charge, eliminating the leading sensitivity of the Higgs potential to $m_*$. While this exceptional \tH model admits a full mirroring of the SM, we find that only when the twin gauge bosons are decoupled above $m_*$, the Higgs VEV and mass can be reproduced with a mild level of fine-tuning, to a good approximation given by $2 \xi \approx 10 \%$. The misalignment $\xi = v^2/f^2 $ between the EW scale and the Higgs decay constant is a common requisite in all the constructions with a composite-NGB Higgs, since its couplings to the SM depart from those of an elementary Higgs proportionally to $\xi$, and such departures are constrained by both EW precision tests and Higgs couplings measurements. This characteristic of \tH models becomes ever more important to test, something that could well be achieved at a future Higgs factory.

The exceptional twin Higgs brings about a different set of thrilling probes characterizing \emph{hypercharged naturalness}. The leading indirect effect is found in the loop-induced Higgs coupling to photons, which receives a contribution from the twin top which makes it more SM-like in comparison to other composite-Higgs models. The fondness of the twins for hypercharge, that is for photons and $Z$'s, shows up in a striking way in processes where the twins are directly produced. In the fraternal scenario, this affinity quickly drives the model into fine-tuned territory, due to the relatively light $Z'$ that couples to the SM quarks and leptons via their mixing with the $B$. Much more compelling is the case where we strip our construction to the bare minimum: the twin pNGBs, the twin top and possibly the twin bottom. While the neutral pseudoscalar $\omega^0$ is almost inert, the electrically charged scalar $\omega^\pm$ is long-lived or stable on collider scales, giving rise to charged tracks that the LHC could soon detect. The twin top/bottom exhibits quirky phenomenology:  once pair-produced they lose energy up until they are brought back together by their twin-QCD interactions, eventually forming low-lying twin-quarkonium states, either a pseudoscalar $\eta^{-+}$ or a vector $\Upsilon^{--}$. The former annihilates mostly to twin gluons, thus becoming another source of twin glueballs besides those from Higgs decays as in other \tH models. The latter either annihilates to SM fermions, thus it could be detected as a $Z'$-like resonance (although with much smaller cross sections than the fraternal $Z'$), or to $\gamma h$ and $Z h$, which can be considered a smoking gun of our exceptional twin Higgs. The production cross sections are small but kinematically accessible at the LHC, and prospects for a detection at the HL-LHC are promising. Finally, the twin glueballs, in particular the lightest $0^{++}$, exhibit similar phenomenology as those in other \tH models: they are long-lived if the twin-QCD scale is not considerably above $\Lqcd$. Amusingly, some of the glueballs such as the $0^{-+}$ decay almost certainly to diphotons, and they live very long, the longer the heavier the twin bottom, a reasonable possibility being this state merely an spectator for what concerns the stability of the electroweak scale. In such a case, these glueballs would leave the LHC detectors as missing energy, and eventually reappear as a pair of photons, to be picked up by a future detector such as MATHUSLA. 

These are experimental opportunities that we believe should not be missed, given the high stakes in the search for the dynamics, even if eerie, that protect us from the unknown at the highest energy scales.\footnote{For the night is dark and full of terrors (Melisandre of Asshai).}


\section*{Acknowledgements}
We thank Ennio Salvioni for useful discussions.
The work of JS, SS, and AW has been partially supported by the DFG Cluster of Excellence 2094 "Origins", by the Collaborative Research Center SFB1258 and BMBF grant no. 05H18WOCA1.


\appendix

\section{$SO(7)$ and $G_2$ representation} \label{embed}

In this appendix we construct the basis of generators for $SO(7)$ in its spinor $\mathbf{8}$ representation.
We start with the $\Gamma$ matrices, here written as outer products of the Pauli matrices,
\bea
\Gamma_1 \!\!\!&=&\!\!\! \sigma_1 \otimes \sigma_0 \otimes \sigma_2 \, , \nn \\
\Gamma_2 \!\!\!&=&\!\!\! \sigma_2 \otimes \sigma_0 \otimes \sigma_0 \, , \nn \\
\Gamma_3 \!\!\!&=&\!\!\! \sigma_1 \otimes \sigma_2 \otimes \sigma_1 \, , \nn \\
\Gamma_4 \!\!\!&=&\!\!\! \sigma_1 \otimes \sigma_2 \otimes \sigma_3 \, , \nn \\
\Gamma_5 \!\!\!&=&\!\!\! \sigma_3 \otimes \sigma_1 \otimes \sigma_2 \, , \nn \\
\Gamma_6 \!\!\!&=&\!\!\! \sigma_3 \otimes \sigma_2 \otimes \sigma_0 \, , \nn \\
\Gamma_7 \!\!\!&=&\!\!\! \sigma_3 \otimes \sigma_3 \otimes \sigma_2 \, .
\label{gamma}
\eea
A basis of generators is then obtained as 
\beq
\label{TM}
M_{ij} = \frac{1}{4 i} \left[ \Gamma_i, \Gamma_j \right] \, .
\eeq
We find convenient to define the linear combinations
\bea
\label{TL}
& T_L^1 = {1 \over 2} \left(M_{14} - M_{23} \right) \, , \,\,\, T_L^2 = {1 \over 2} \left(M_{13} + M_{24} \right) \, , \,\,\, T_L^3 = {1 \over 2} \left(M_{12} - M_{34} \right) \, , & \\
\label{TLt}
& T_{\Lt}^1 = {1 \over 2} \left(M_{14} + M_{23} \right) \, , \,\,\, T_{\Lt}^2 = {1 \over 2} \left(M_{13} - M_{24} \right) \, , \,\,\, T_{\Lt}^3 = - {1 \over 2} \left(M_{12} + M_{34} \right) \, , & \\
\label{TZ}
& T_{\Z}^1 = M_{67} \, , \,\,\, T_{\Z}^2 = - M_{57} \, , \,\,\, T_{\Z}^3 = M_{56} \, , &
\eea
which we identify with the generators of the $SU(2)_L \times SU(2)_{\Lt} \times SU(2)_{\Z}$ subgroups of $SO(7)$. We have normalized the generators as $\Tr[T^A T^B] = \delta^{AB}$, except for those generating $SU(2)_{\Z}$, which again for convenience we define with norm $\Tr[T_{\Z}^i T_{\Z}^j] = 2 \delta^{ij}$.

The unbroken $SU(2)_{R = \Lt + \Z}$ subgroup in $G_2$ and the corresponding axial combination in $SO(7)/G_2$ are identified as
\bea
\label{TR}
& T_R^{i} = T_{\Lt}^i + T_{\Z}^i \, , & \\
\label{THt}
& 
T^{\hat a = i+4} = {1 \over \sqrt{3}} \big( T_{\Lt}^i - T_{\Z}^i \big) \, , &
\eea
with $\Tr[T_{R}^i T_{R}^j] = 3 \delta^{ij}$.
The Higgs generators are
\bea
& 
T^{\hat a = 2} = - {1 \over \sqrt{6}} \left(M_{15} - M_{26} - M_{37} \right) \, , \,\,\, 
T^{\hat a = 1} = {1 \over \sqrt{6}} \left(M_{16} + M_{25} + M_{47} \right) \, , & \nn \\ 
& 
T^{\hat a = 4} = -{1 \over \sqrt{6}} \left(M_{17} + M_{35} - M_{46} \right) \, , \,\,\, 
T^{\hat a = 3} = {1 \over \sqrt{6}} \left(M_{27} - M_{36} - M_{45} \right) \, ,&
\label{TH}
\eea
while the rest of unbroken generators, in $G_2$, are
\bea
\label{THp}
& 
{1 \over 2} \left(M_{15} + M_{26} \right) \, , \,\,\, 
{1 \over 2} \left(M_{16} - M_{25} \right) \, , \,\,\, 
{1 \over 2} \left(M_{35} + M_{46} \right) \, , \,\,\, 
{1 \over 2} \left(M_{36} - M_{45} \right) \, , & \\
\label{THpp}
& 
{1 \over 2 \sqrt{3}} \left(M_{15} - M_{26} + 2 M_{37} \right) \, , \,\,\ 
{1 \over 2 \sqrt{3}} \left(M_{16} + M_{25} - 2 M_{47} \right) \, , & \nn \\ 
& 
{1 \over 2 \sqrt{3}} \left(2 M_{17} - M_{35} + M_{46} \right) \, , \,\,\,
- {1 \over 2 \sqrt{3}} \left(2 M_{27} + M_{36} + M_{45} \right) \, .&
\eea

\bibliographystyle{mine}
\bibliography{exceptional}

\end{document}